\documentclass{emulateapj}
\usepackage[backref,breaklinks,colorlinks,citecolor=blue]{hyperref}
\usepackage[all]{hypcap}

\usepackage{amsmath}

\newcommand{\myemail}{adams@psi.edu}
\newcommand{\kepler}{\emph{Kepler}}
\newcommand{\ktwo}{\emph{K2}}
\newcommand{\epicexample}{EPIC 201637175}
\newcommand{\nb}{100}
\newcommand{\numtrialperiods}{10,000}
\newcommand{\qmi}{1\%}
\newcommand{\qma}{50\%}
\newcommand{\thrustersignalperiod}{5.88460}

\newcommand{\czerotargetnum}{7,746}
\newcommand{\conetargetnum}{21,647}
\newcommand{\ctwotargetnum}{13,401}
\newcommand{\cthreetargetnum}{16,375}
\newcommand{\cfivetargetnum}{15,779}
\newcommand{\cfourtargetnum}{24,475}

\newcommand{\ebnum}{91}                             

\newcommand{\eeblssnrthresh}{10}
\newcommand{\eeblssnrthreshnumzero}{757}   
\newcommand{\eeblssnrthreshnumone}{505}    
\newcommand{\eeblssnrthreshnumtwo}{1251}   
\newcommand{\eeblssnrthreshnumthree}{419}    
\newcommand{\eeblssnrthreshnumfour}{1480}   
\newcommand{\eeblssnrthreshnumfive}{1032}   

\newcommand{\durationthreshnumzero}{91}      
\newcommand{\durationthreshnumone}{42}      
\newcommand{\durationthreshnumtwo}{177}      
\newcommand{\durationthreshnumthree}{47}   
\newcommand{\durationthreshnumfour}{152}     
\newcommand{\durationthreshnumfive}{134}      

\newcommand{\totalsnr}{99423}             
\newcommand{\totalsnrdurdepth}{5444} 
\newcommand{\totalcand}{19}                
\newcommand{\dodgycand}{4}                

\newcommand{\candidateszero}{1}         
\newcommand{\candidatesone}{4}         
\newcommand{\candidatestwo}{2}         
\newcommand{\candidatesthree}{5}         
\newcommand{\candidatesfour}{7}         
\newcommand{\candidatesfive}{4}         

\slugcomment{Accepted to AJ on 2016 May 20}

\usepackage{aas_macros}
\usepackage{graphics}
\usepackage{amsmath}
\usepackage{natbib}
\begin{document}


\title{Ultra Short Period Planets in \ktwo\: SuPerPiG Results for Campaigns 0-5}


\author{Elisabeth R. Adams}
\affil{Planetary Science Institute, 1700 E. Ft. Lowell, Suite 106, Tucson, AZ 85719, USA}
\email{\myemail}

\author{Brian Jackson}
\affil{Department of Physics, Boise State University, 1910 University Drive, Boise ID 83725, USA}



\author{Michael Endl}
\affil{McDonald Observatory, The University of Texas at Austin, Austin, TX 78712, USA}






\begin{abstract}

We have analyzed data from Campaigns 0-5 of the \ktwo\ mission and report \totalcand\ ultra-short-period candidate planets with orbital periods of less than 1 day (nine of which have not been previously reported). Planet candidates range in size from 0.7-16 Earth radii and in orbital period from 4.2 to 23.5 hours. One candidate (EPIC 203533312, Kp=12.5) is among the shortest-period planet candidates discovered to date ($P=4.2$~hours), and, if confirmed as a planet, must have a density of at least  $\rho=8.9\ {\rm g/cm^3}$ in order to not be tidally disrupted. Five candidates have nominal radius values in the sub-Jovian desert ($R_P = 3-11~R_{\oplus}$ and $P \le 1.5$~days) where theoretical models do not favor their long-term stability; the only confirmed planet in this range is in fact thought to be disintegrating (EPIC 201637175). In addition to the planet candidates, we report on four objects which may not be planetary, including one with intermittent transits (EPIC 211152484) and three initially promising candidates that are likely false positives based on characteristics of their light curves and on radial velocity follow-up. A list of \ebnum\ suspected eclipsing binaries identified at various stages in our vetting process is also provided. Based on an assessment of our survey's completeness, we estimate an occurrence rate for ultra-short period planets among \ktwo\ target stars that is about half that estimated from the {\it Kepler} sample, raising questions as to whether \ktwo\ systems are intrinsically different from {\it Kepler} systems, possibly as a result of their different galactic location.

\end{abstract}



\section{Introduction}

Planets with orbital periods of less than a day present real challenges to theories of planet formation and evolution, and yet numerous objects with periods as short as a few hours have been found. So close to their host stars that some are actively disintegrating \citep{2015ApJ...812..112S}, these planets' origins remain unclear, and even modified models for planet formation and evolution with significant inward migration have trouble accounting for their periods. They thus present an important test for theories of planetary origins and evolution. 

The existence of such a population was suggested by several groups. For instance, \citet{2009ApJ...698.1357J}, among others, suggested that orbital decay driven by tides raised on the host star could drive planets into very short-period orbits, but how they arrived within orbits susceptible to tidal decay is unclear. \citet{2008MNRAS.384..663R} explored different possibilities, from in situ formation to Type-1 gas disk migration. That study suggested the orbital architectures of the systems, along with the physical properties of the planets, could help distinguish between the possible origin scenarios. For instance, Type-1 migration would be expected to move multiple, small planets together into short-period orbits with the planets forming a chain of mean-motion (or near mean-motion) resonances (MMRs). The magnetosphere of a young star is thought to clear a cavity within a few stellar radii of the star, and so the inward migration is expected to cease shortly after entering that cavity, depositing a migrating planet into an orbital period of a few days \citep{1996Natur.380..606L}. 

Among the discoveries reported in \citet{SanchisOjeda2014} are several multi-planet systems, and in some cases, the outer pairs of planets are near MMRs. For example, the KOI-1843 system includes two outer planets, with $P \approx 4$ and 6 days, within 1\% of the 3:2 MMR. The innermost planet, KOI-1843.03 has an orbital period $P = 4$ hours, more than 75 Hill radii from its nearest sibling, and so presumably is dynamically decoupled. However, tidal decay could potentially explain KOI-1843.03's present precarious orbit, especially if interactions with the other two planets excited its orbital eccentricity and enhancing driving tidal decay \citep{2014PhDT.........4V}. Indeed, KOI-1843.03 is so close to its host star ($\approx 2$ stellar radii) that the planet currently orbits within space originally occupied by its host star early in the system's history. Thus, it must have arrived at its present orbit long after the system's formation. Many of the ultra-short-period (USP) planet and candidates occupy similar orbits and therefore tidal damping seems likely to have played a role in shaping their orbits. 

Whatever the planets' origins, the properties of the host stars seem related to the planets' occurrence rate. \citet{SanchisOjeda2014} found marginal evidence that M-dwarfs are about 10 times as likely as F-dwarfs to host a planet with a radius twice that of Earth's and $P \le$ 24 hours. This trend is qualitatively similar to but more pronounced than the trend discussed in \citet{2012ApJS..201...15H} but disagrees with that discussed in \citet{2013ApJ...766...81F}, who found no dependence of the occurrence rate on stellar type; however, neither study considered USP planets. 

Presumably, the properties of the host stars play a role in the planet formation process itself but probably also shape evolution of the planetary system. For instance, if stars of all types were equally likely to host a planet, tidal decay of a planet's orbit should occur more quickly for more massive, larger stars. The influence of the tide raised on the star rapidly increases with stellar radius, and the influence of the tide raised on the planet scales with stellar mass. On the other hand, if Type-1 migration is, indeed, responsible for bringing small planets close-in, then planets orbiting M-dwarfs might start their tidal journey closer to the star, and therefore be more susceptible to tidal decay.

Unfortunately, the small sample size of USP planetary candidates reported in \citet{SanchisOjeda2014} around M-dwarfs (six) and F-dwarfs (nine) makes it difficult to draw statistically robust inferences regarding the dependences of USP planet occurrence rate. Thankfully, the reincarnation of the {\it Kepler} Mission as \ktwo\ provides an ideal opportunity to continue the search for USP planets. Although \ktwo\'s 80-day dwell time on each target field makes finding longer period planet difficult, USPs are easily detectable as a transiting planet with $P =$ 4 hours occults its host star almost 480 times in 80 days. 

In this paper we report on the ongoing efforts of the Short-Period Planets Group (SuPerPiG, \url{http://www.astrojack.com/research/superpig/}) to find additional short-period planets in the \ktwo\ mission data, using the light curve products produced by the k2sff pipeline of \citet{2014PASP..126..948V}. In Section \ref{section:data_conditioning_and_transit_search}, we detail how  we processed the \ktwo\ data and our transit search and model-fitting. In Section \ref{section:cands}, we present our candidates, discuss their occurrence rate and the properties of their host stars. We also quantify our survey's completeness. In Section \ref{section:discussion}, we discuss future prospects and the possibilities for follow-up observations.

\section{Data Conditioning and Transit Search}
\label{section:data_conditioning_and_transit_search}

\begin{figure}
\includegraphics[width=0.5\textwidth]{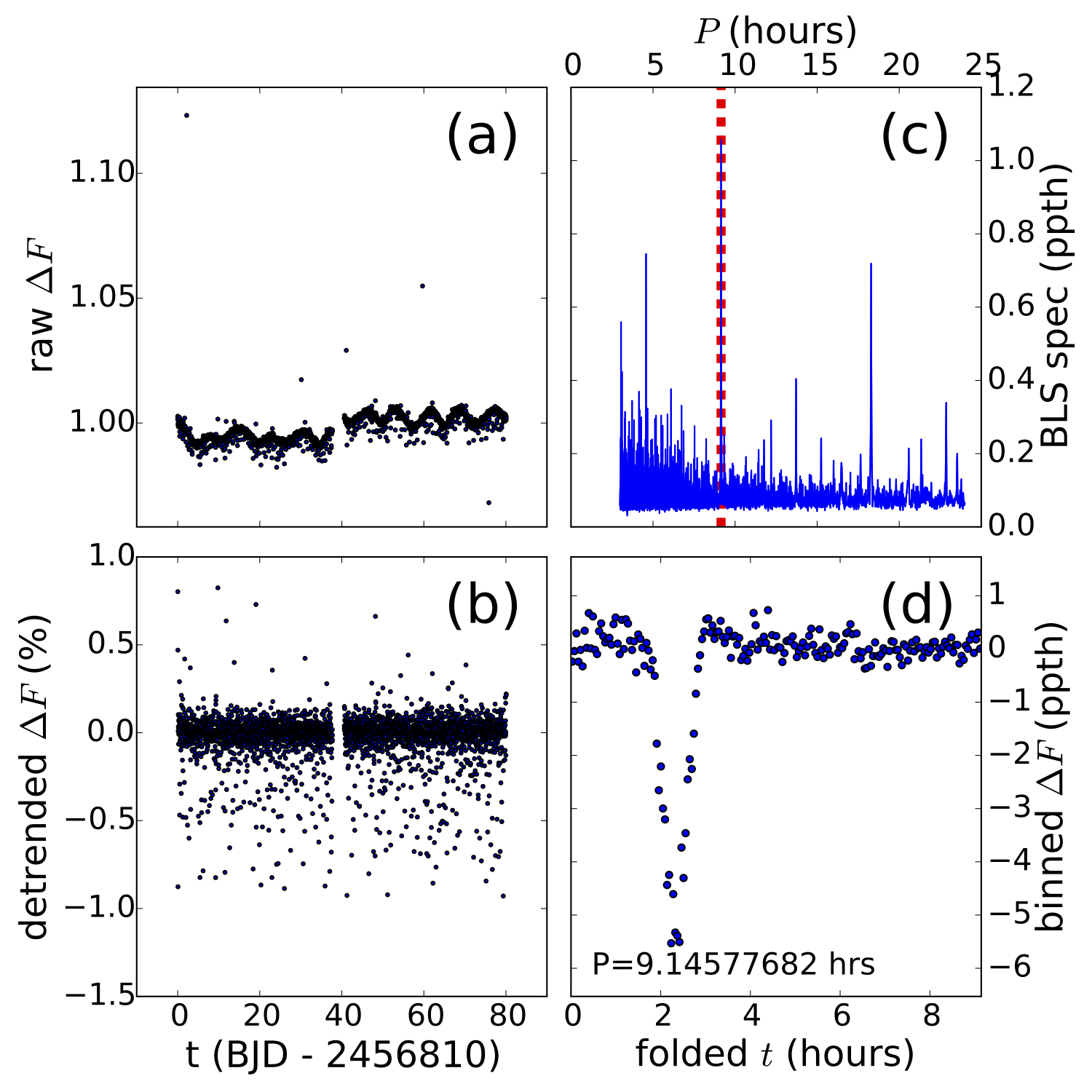}
\caption{Our data conditioning and transit search process as applied to the \epicexample\ dataset from \ktwo\ Campaign 1, which was first identified by \citet{2015ApJ...812..112S}. (a) The raw k2sff data. (b) The k2sff data after applying a median boxcar filter with width of 1-day. (c) The resulting EEBLS spectrum with the best-fit period hightlighted by a dashed, red line. (d) The detrended data from (b), folded on the best-fit period and binned into 200 bins for display. Note (a) and (b) share x-axes, but (c) and (d) do not.}
\label{fig:example_transit}
\end{figure}

\begin{figure}
\includegraphics[width=0.5\textwidth]{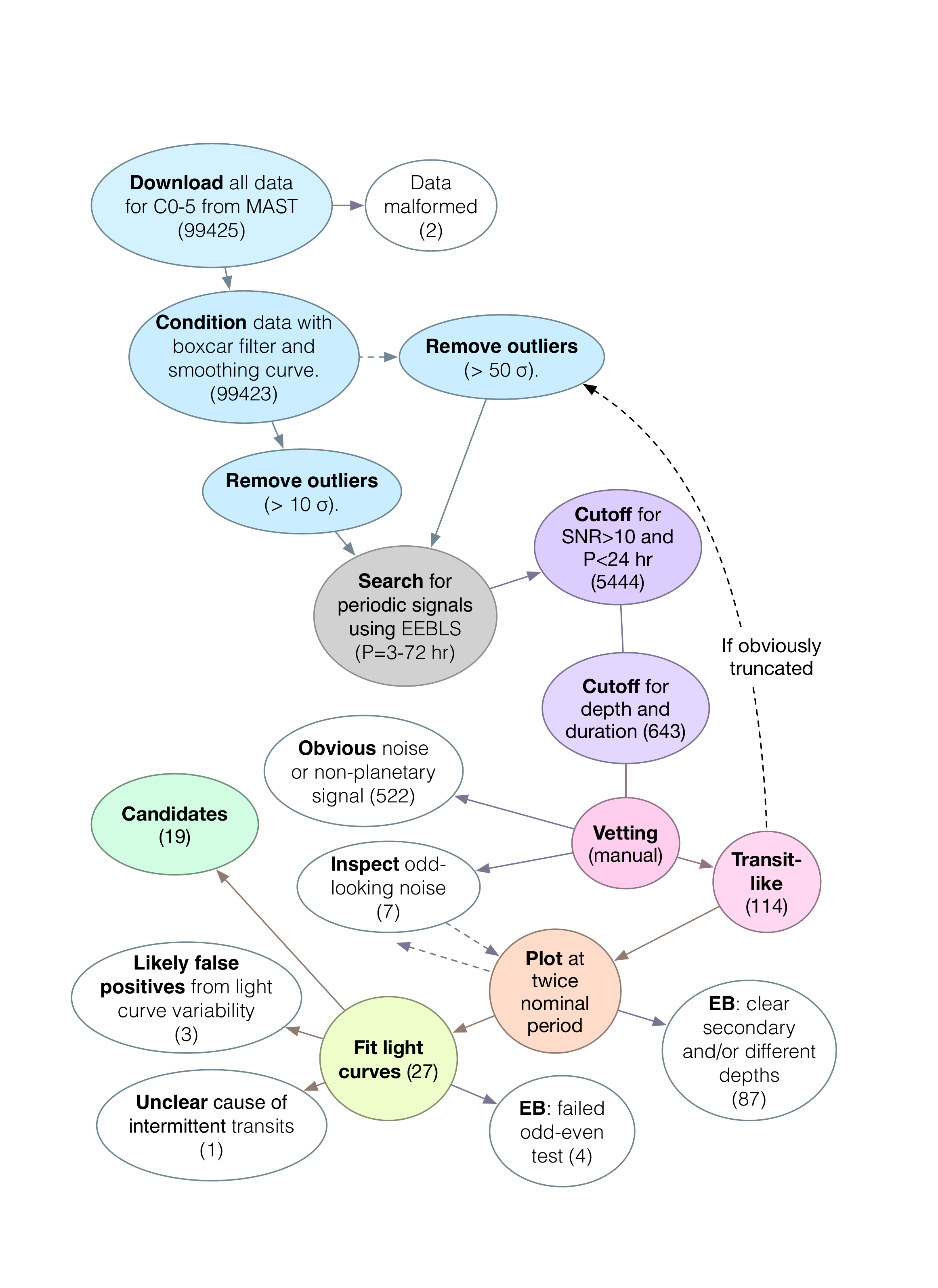}
\caption{Flowchart of the transit candidate search and vetting process.}
\label{fig:flowchart}
\end{figure}

The process applied here follows closely that in \citet{2013ApJ...779..165J}, and we verified its efficacy by recovering the disintegrating planet EPIC 201637175b reported in \citet{2015ApJ...812..112S}. \autoref{fig:example_transit} illustrates our data conditioning and transit search process as applied to that dataset. We also recovered synthetic transits that we injected into the data to test our survey's completeness, as described in Section \ref{section:survey_completeness}. The complete flowchart of our search and vetting process is shown in \autoref{fig:flowchart}.

We retrieved the \anchor{\url{https://archive.stsci.edu/prepds/k2sff/}}{k2sff publicly available data} generated by the pipeline described in \citet{2014PASP..126..948V} for all \totalsnr targets from the MAST archive for campaigns 0 to 5, C0-C5. (Two objects in C0, 202093417 and 202137146, were not used because the flux returned from MAST was uniformly -1.) The search includes targets of all object type, although only one ultimate candidate (202094740) is listed as ``None'' rather than ``STAR'' in MAST. An advantage to using the \citet{2014PASP..126..948V} pipeline is that it corrects for the pointing drift of the \ktwo\ spacecraft and generates light curves for several different photometric apertures, with a ``best aperture'' corresponding to the aperture giving the smallest RMS variation in the final light curve. Only long-cadence data, with a 30 minute exposure time, was searched. For our initial search, we used the best aperture light curves (the ``BESTAPER'' extension in the FITS files) for the targets available (\czerotargetnum\ in C0, \conetargetnum\ in C1, \ctwotargetnum\ in C2,  \cthreetargetnum\ in C3, \cfourtargetnum\  in C4, and \cfivetargetnum\ in C5). The processed data for \epicexample\ from C1 \citep[previously discovered by ][]{2015ApJ...812..112S} are shown in \autoref{fig:example_transit} (a).

\begin{figure}
\includegraphics[width=0.5\textwidth]{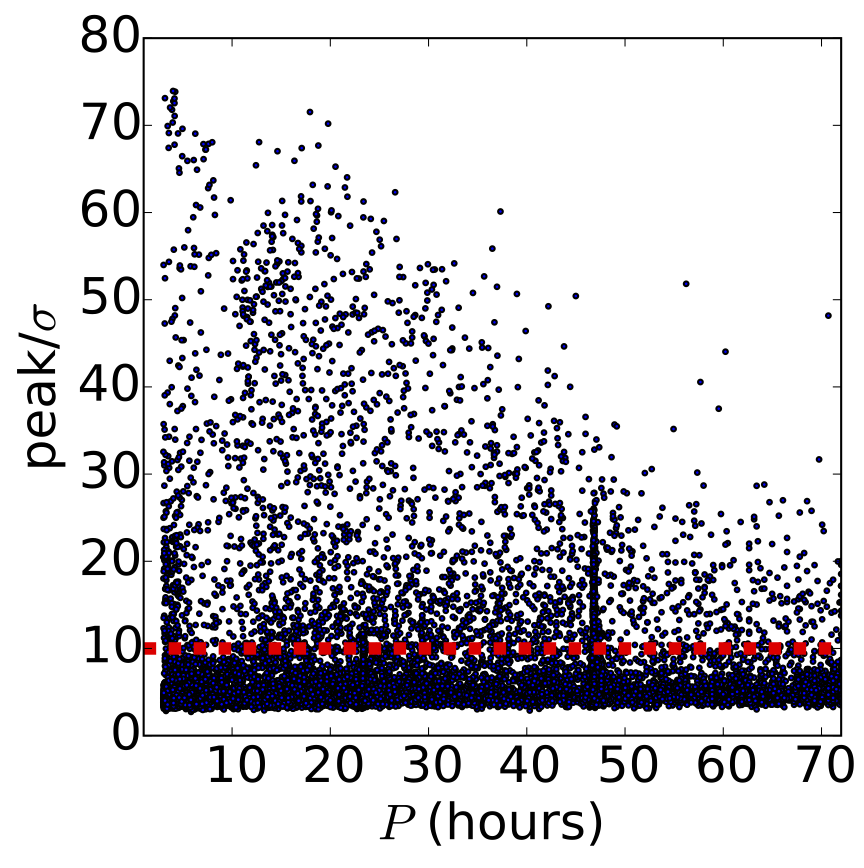}
\caption{Distribution of best-fit transit periods (3-72 hours) from EEBLS, shown for Campaign 2, with the strength of the EEBLS signal at the peak as scaled by each spectrum's $\sigma$. The red dashed line shows the 10$\sigma$ cutoff used to identify candidates. Only signals between 3-24 hours were the subject of this analysis. Residual effects of the thruster firing period can be seen in an artificial pileup at 47 hours (8x), and spurious signals, which don't resemble actual transit light curves, remain in all campaigns at multiples (typically 1x and 8x) of the thruster period.}
\label{fig:final_eebls_periods_c02}
\end{figure}

These light curves still exhibit a variety of astrophysical and instrumental variations that act as sources of noise for our analysis, but most of that variability is on timescales longer than the periods of interest for our search. To mitigate these variations, we applied a median boxcar filter with a width of 1-day. The original data have an almost regular observing cadence of 30-min with a few small gaps, so we first linearly interpolated the light curves to a grid with a completely regular sampling of 30-min and then calculated the median value for all points within an 1-day window of each regularly-gridded point to generate our filter. We then interpolated this filter back to the time grid of the original data, restoring the original time gaps, and subtracted this interpolated filter from the original data. To improve detection efficiency for the shallowest transits, we masked out the handful of data points lying more than 10-$\sigma$ from the dataset median, where the standard deviation $\sigma \equiv 1.4826\ \times$ the median absolute deviation \citep{Leys2013764}. (For deeper transits where a 10-$\sigma$ cutoff removes real data from in transits, the truncated transits are still easily detected, and the lightcurves were regenerated with the noise filter set to 50-$\sigma$ before fitting to get the correct transit parameters.) The resulting detrended data for \epicexample\ are shown in \autoref{fig:example_transit} (b).

\capstartfalse
\begin{deluxetable*}{llll  |  llll   | llll}
\tablewidth{0pt}
\tablecaption{Eclipsing binaries with original EEBLS periods from 3-24 hours}
\tablehead{
\colhead{EB} & \colhead{Campaign} & \colhead{Period} &\colhead{Notes} &\colhead{EB} & \colhead{Campaign} & \colhead{Period} &\colhead{Notes} &\colhead{EB} & \colhead{Campaign} & \colhead{Period} & \colhead{Notes}\\
\colhead{} & \colhead{} & \colhead{days} &\colhead{} & \colhead{} & \colhead{} & \colhead{days} &\colhead{} & \colhead{} & \colhead{} & \colhead{days} &\colhead{}
}
\startdata
202073210	&C0	&1.024979	&AB	&202083924	&C0	&0.37845		&AB	&202087156	&C0	&1.893746	&A\\
202088191	&C0	&1.662009	&AB	&202091545	&C0	&1.857576	&AB	&202103762	&C0	&1.327711	&AB\\
201182911	&C1	&1.993038	&AB	&201184068 	&C1	&1.588528	&AB	&201523873 	&C1	&1.240377	&AB\\
201563164 	&C1	&0.37491		&AC	&201607088 	&C1	&0.53093		&B	&201649211 	&C1	&0.199731	&\\
201680569 	&C1	&0.784835	&AB	&201691826 	&C1	&0.899402	&AB	&201740472 	&C1	&0.958733	&\\
201810513 	&C1	&1.646639	&AB	&201843069 	&C1	&1.096823	&AB	&201848566 	&C1	&0.956863	&AD\\
201893576 	&C1	&0.929699	&AB	&201903318 	&C1	&0.390019	&A	&202828096	&C2	&1.436287	&AB\\
202971774 	&C2	&0.471778	&A	&203027459 	&C2	&0.530881	&	&203633064	&C2	&0.709932	&B\\
204429688 	&C2	&0.410635	&	&204470067 	&C2	&1.846896	&A	&204538608 	&C2	&0.914888	&AB\\
204822463 	&C2	&1.208611	&A	&205068000 	&C2	&0.686159	&	&205129673 	&C2	&0.955311	&B\\
205377483 	&C2	&0.787653	&A	&205899208	&C3	&0.289449	&A	&205910324 	&C3	&0.25624		&A\\
205934874 	&C3	&1.236054	&A	&205962262 	&C3	&0.849755	&B	&205978103 	&C3	&0.64287		&AB\\
205996447 	&C3	&1.600526	&AB	&206045146 	&C3	&0.233232	&	&206050740 	&C3	&0.198387	&\\
206100943 	&C3	&1.601195	&AB	&206109113 	&C3	&1.323701	&AB	&206139574 	&C3	&0.673044	&B\\
206315178 	&C3	&0.635412	&A	&206489474 	&C3	&1.516983	&AB	&210404228	&C4	&1.119943	&AB\\
210434247 	&C4	&0.453837	&B	&210574135 	&C4	&0.929409	&B	&210593417 	&C4	&1.062117	&AB\\
210659779 	&C4	&0.469356	&A	&210662654 	&C4	&0.159109	&B	&210663545 	&C4	&0.300752	&A\\
210664740 	&C4	&0.414948	&	&210675130 	&C4	&1.377904	&AB	&210754505 	&C4	&1.741556	&A\\
210821360 	&C4	&1.467197	&AD	&210843708 	&C4	&0.704035	&	&210863062 	&C4	&0.644008	&B\\
210932768 	&C4	&0.151526	&	&210941737 	&C4	&1.154352	&AB	&210954667 	&C4	&0.325575	&A\\
211012889 	&C4	&1.7392		&A	&211315506	&C5	&0.884559	&	&211380136 	&C5	&1.753621	&AB\\
211389268 	&C5	&0.28499		&AB	&211518347 	&C5	&1.871362	&AB	&211526186 	&C5	&0.455062	&B\\
211578677 	&C5	&0.554273	&AB	&211580526 	&C5	&1.758335	&A	&211604668 	&C5	&0.515911	&B\\
211613886 	&C5	&0.958809	&AB	&211623903 	&C5	&1.615615	&AB	&211631904 	&C5	&0.442015	&B\\
211685048 	&C5	&0.769125	&B	&211719362 	&C5	&0.402237	&AB	&211719484 	&C5	&0.721766	&AB\\
211796803 	&C5	&0.277589	&A	&211797674 	&C5	&1.323035	&A	&211822953 	&C5	&1.549424	&AB\\
211833449 	&C5	&0.534773	&AB	&211833616 	&C5	&0.534745	&AB	&211906940 	&C5	&0.17545		&\\
211931594 	&C5	&0.320086	&AD	&211953866 	&C5	&1.788937	&AB	&211978865 	&C5	&0.907764	&B\\
211995966 	&C5	&0.558524	&AB	&211999656 	&C5	&1.948462	&AB	&212066407 	&C5	&1.643616	&AD\\
212069706 	&C5	&0.273843	&A	&212083250 	&C5	&0.518772	&B	&212155299 	&C5	&0.901706	&B\\
212158225 	&C5	&0.266589	&A	&	&	&	&	&	&	&	&
\enddata
\tablecomments{A: Initial EEBLS period was half the period listed here. B: Object is listed in Kepler EB catalog (Third Revision Beta) \url{http://keplerebs.villanova.edu/}. C: Identified as a disintegrating minor planet around a white dwarf by \citet{2015Natur.526..546V}. D: Object is listed at half correct period in Kepler EB catalog (Third Revision Beta) \url{http://keplerebs.villanova.edu/}. }
\label{table:eb}
\end{deluxetable*}
\capstarttrue

\capstartfalse
\begin{deluxetable*}{cc cc ll ll ll ll}
\tablewidth{0pt}
\tablecaption{Parameters for Stars Hosting Candidate Planets}
\tablehead{
\colhead{Candidate} & \colhead{Campaign} & \colhead{RA} & \colhead{Dec} & \colhead{Kp} & \colhead{$R_\star$ ($R_\odot$)} & \colhead{$T_{\rm eff}$ (K)} & \colhead{[Fe/H]} & \colhead{$\log(g)$} & \colhead{$u_1$} & \colhead{$u_2$} & \colhead{Source} 
}
\startdata
$202094740$	&C0	&$100.46312$		&$27.0972$	&$11.5$	&$1.08\pm0.04$	&$6481.0$	&$0.0$	&$4.0$	&$0.297$		&$0.3126$	&TESS-Vanderbilt\\
$201264302$	&C1	&$169.598013$	&$-2.991577$	&$13.88$	&$0.26\pm0.05$	&$3299.0$	&$0.155$	&$5.106$	&$0.4168$	&$0.3719$	&EPIC\\
$201606542$	&C1	&$170.311881$		&$2.144426$	&$11.92$	&$0.82\pm0.05$	&$5540.0$	&$0.03$	&$4.81$	&$0.4843$	&$0.1926$	&McDonald, A\\
$201637175$	&C1	&$169.482818$	&$2.61907$	&$14.93$	&$0.54\pm0.07$	&$3830.0$	&$0.03$	&$4.65$	&$0.5976$	&$0.157$		&S15b\\
$201650711$	&C1	&$172.044052$	&$2.826891$	&$12.25$	&$0.69\pm0.05$	&$4340.0$	&$-0.81$	&$4.25$	&$0.5126$	&$0.2191$	&McDonald, B\\
$203533312$	&C2	&$243.955378$	&$-25.818471$	&$12.16$	&$1.15\pm0.08$	&$6620.0$	&$0.09$	&$4.19$	&$0.2967$	&$0.3076$	&McDonald\\
$205152172$	&C2	&$245.180008$	&$-19.14414$	&$13.49$	&$0.66\pm0.1$		&$4202.0$	&$-0.04$	&$4.758$	&$0.7146$	&$0.0666$	&EPIC\\
$206103150$	&C3	&$331.203044$	&$-12.018893$	&$11.76$	&$1.15\pm0.04$	&$5565.0$	&$0.43$	&$4.29$	&$0.5247$	&$0.1718$	&S15a\\
$206151047$	&C3	&$333.71501$		&$-10.769168$	&$13.43$	&$0.89\pm0.06$	&$5928.0$	&$-0.169$	&$4.247$	&$0.3879$	&$0.2573$	&EPIC\\
$206169375$	&C3	&$344.016383$	&$-10.332234$	&$12.56$	&$0.95\pm0.07$	&$6167.0$	&$-0.168$	&$4.16$	&$0.3258$	&$0.2987$	&EPIC\\
$206298289$	&C3	&$337.357688$	&$-8.266702$	&$14.69$	&$0.5\pm0.06$		&$3724.0$	&$0.017$	&$4.942$	&$0.355$		&$0.3581$	&EPIC\\
$206417197$	&C3	&$337.133466$	&$-6.347505$	&$13.35$	&$0.77\pm0.04$	&$5007.0$	&$-0.063$	&$4.587$	&$0.5711$	&$0.1407$	&EPIC\\
$210414957$	&C4	&$64.325266$		&$13.804842$	&$12.65$	&$0.87\pm0.05$	&$5838.0$	&$0.3$	&$4.12$	&$0.4643$	&$0.2162$	&McDonald\\
$210754505$	&C4	&$64.64493$		&$19.179056$	&$13.19$	&$0.88\pm0.05$	&$5875.0$	&$0.04$	&$4.04$	&$0.4326$	&$0.2302$	&McDonald\\
$210605073$	&C4	&$65.62329$		&$17.037875$	&$17.89$	&$1.42\pm0.54$	&$7020.0$	&$0.0$	&$4.0$	&$0.2948$	&$0.2825$	&Photometry\\
$210707130$	&C4	&$59.464394$		&$18.465254$	&$12.1$	&$0.71\pm0.05$	&$4462.0$	&$-0.28$	&$4.17$	&$0.6686$	&$0.0897$	&McDonald\\
$210954046$	&C4	&$60.903766$		&$22.249157$	&$12.44$	&$0.94\pm0.1$		&$6125.0$	&$0.0$	&$3.0$	&$0.3583$	&$0.2623$	&McDonald, C\\
$210961508$	&C4	&$59.920104$		&$22.365984$	&$13.56$	&$0.76\pm0.04$	&$4925.0$	&$-0.06$	&$3.42$	&$0.5599$	&$0.1535$	&McDonald\\
$211152484$	&C4	&$60.069895$		&$25.48$		&$12.14$	&$0.96\pm0.05$	&$6188.0$	&$-0.12$	&$4.12$	&$0.341$		&$0.2782$	&McDonald\\
$211357309$	&C5	&$133.23263$		&$10.944721$	&$13.15$	&$0.42\pm0.06$	&$3563.0$	&$0.097$	&$5.004$	&$0.3986$	&$0.3402$	&EPIC\\
$211685045$	&C5	&$123.359621$	&$15.713759$	&$14.98$	&$0.54\pm0.13$	&$3832.0$	&$-0.058$	&$4.9$	&$0.4294$	&$0.2983$	&EPIC\\
$211995325$	&C5	&$131.207598$	&$20.177694$	&$18.22$	&$0.62\pm0.22$	&$4071.0$	&$-0.202$	&$4.838$	&$0.4525$	&$0.259$		&EPIC\\
$212150006$	&C5	&$128.169544$	&$23.031999$	&$14.7$	&$0.82\pm0.05$	&$5528.0$	&$-0.281$	&$4.482$	&$0.4385$	&$0.23$		&EPIC
\enddata
\tablecomments{$T_{\rm eff}$, $\log(g)$, and [Fe/H] are taken from the source listed; EPIC 202094740 and 210605073 assumed  $\log(g)=4.0$ and $[Fe/H]=0$. Quadratic limb darkening parameters $u_1$ and $u_2$ are derived from \citet{2011A&A...529A..75C}.
A: Double-peaked spectra (planetary parameters are strongly diluted, possible false positive).
B. McDonald guide camera companion at 1\arcsec\ and a few magnitudes fainter (planetary parameters are diluted, possible false positive).
C: McDonald guide camera companion at 2\arcsec; also strong nightly RV variation (confirmed false positive).
S15b: \cite{2015ApJ...812..112S}
S15a: \cite{2015ApJ...812L..11S} }
\label{table:stars}
\end{deluxetable*}
\capstarttrue

Using these data, we searched for short-period planetary transits with the EEBLS algorithm \citep{2002A&A...391..369K}. Briefly, the EEBLS algorithm folds and bins data on trial orbital periods and, considering all relevant phases for a given period, returns the best-fit to a square wave at that period. (The original EEBLS algorithm report best-fit periods correspond to positive or negative square waves, i.e. it doesn't care whether the best-fit signal is a dip or a blip. We modified the algorithm slightly to report best-fit periods only for dips.) We started searching for transit signals with periods $P$ between 3 hrs (an orbit near the surface of the Sun) and 3 days (although we are only interested in periods of 1 day or less, the longer initial search threshold is useful to avoid aliases). We used \numtrialperiods\ trial periods, \nb\ bins into which data folded on a trial period were binned, and sought transits with durations between \qmi\ and \qma\ of the trial orbital period. With so many trial periods, the results at one period are not linearly independent from another \citep[cf.][]{2002A&A...391..369K}, but that consideration does not affect whether we recover a transit. Moreover, given the short durations expected ($\lesssim 1$ hour), such a fine gridding is required to recover transits. This analysis produced an EEBLS spectrum for each target, an example of which is shown for \epicexample\ in \autoref{fig:example_transit} (c) with a clear peak at $P \approx 9$ hours.

\autoref{fig:final_eebls_periods_c02} shows the best-fit periods for candidate transit signals from all \ctwotargetnum\ EEBLS spectra in C2 (other campaigns are similar) and the strength of the EEBLS signal at the peak as scaled by each spectrum's $\sigma$. The thruster on-board the \kepler\ spacecraft fires to maintain pointing as needed, with firings occurring at multiples of $P_{\rm thruster} = \thrustersignalperiod$ hours. Mindful of the problems that the thruster firing causes in identifying short period planets \citep{2014PASP..126..948V}, we have masked it out by removing points flagged `MOVING' in the k2sff data. However, in all campaigns even after masking the thruster points, there remains a strong peak at 47 hours (or 8 times the thruster period), as well as a peak at the thruster period itself. These spurious signals do not pass into the final candidate pool, however, since they do not pass the SNR and duration threshholds.

\begin{figure}
\includegraphics[width=0.5\textwidth]{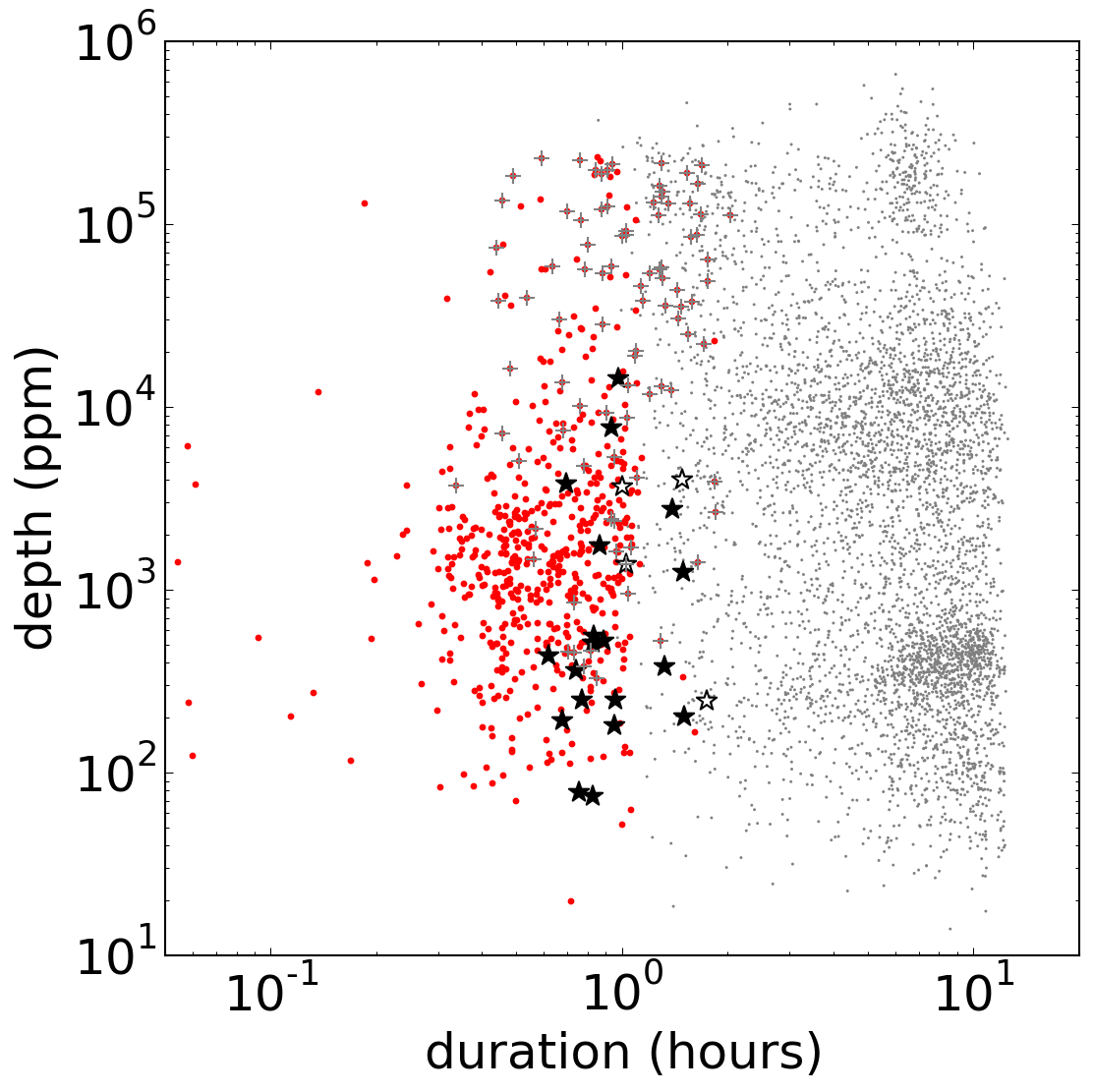}
\caption{Transit durations $\tau$ and depths $D$ (in parts-per-million, ppm) from EEBLS search. All \totalsnr\ detections from C0-5 above the SNR threshhold are plotted in small gray points, while the \totalsnrdurdepth\ larger red points are those with plausible-enough depths and durations relative to the period to warrant further inspection. The \totalcand\ planet candidates that survived subsequent scrutiny are marked by black stars, with the \dodgycand\ suspect candidates as white stars. For comparison, the \ebnum\ eclipsing binaries we identified are shown as grey pluses, and generally, though not exclusively, cluster toward deeper depths.}
\label{fig:duration-depth_threshold}
\end{figure}

To identify candidate transiting planets, we focused on those with EEBLS spectral peaks $\ge$ \eeblssnrthresh-$\sigma$ and periods less than one day, giving  \eeblssnrthreshnumzero,  \eeblssnrthreshnumone, \eeblssnrthreshnumtwo, \eeblssnrthreshnumthree,  \eeblssnrthreshnumfour, and \eeblssnrthreshnumfive\ candidates in C0-C5, respectively, as shown in \autoref{fig:duration-depth_threshold}. We then applied a generous first-pass depth requirement, requiring transit depths $D \le 0.25$, which corresponds to the largest known planet ($2~R_{J}$) transiting a $0.4~R_{\odot}$ M-dwarf. All initial detections with depths from 0.02-0.25 (about 10 per campaign) proved on inspection to be either clear eclipsing binaries or else misidentified stellar noise. We also required durations $\tau$ short enough to be consistent with FGK host star main-sequence densities $\rho_*$ \citep{2003ApJ...585.1038S}: 
\begin{align}
\sin &\left[ \dfrac{\pi \left( \tau - {\rm 1\ hour} \right)}{P} \right] \le \\ 
&\left( \dfrac{3}{4 \pi} \right)^{1/3} \left( R_{\odot} / 1 {\rm AU} \right) \left( P / 1 {\rm yr} \right)^{-2/3} \left(  \rho_*/\rho_\odot \right)^{-1/3} \nonumber \\
& = \Gamma  \left( P / 1 {\rm yr} \right)^{-2/3},
\end{align}
with $\Gamma = 0.00363$ \citep{2013ApJ...779..165J}. An extra hour (or two \ktwo\ long cadence exposure times) was added to the maximum duration ($\tau$) to account for the distortion introduced into the light curves by \ktwo's 30-min observing cadence, which artificially extends the apparent transit duration and makes the transits more V-shaped than is typical for planetary transits. {\bf This threshold gives rise to an apparent cut-off in durations in Figure \ref{fig:duration-depth_threshold} longward of $\sim$ 1-hour. The origin of other patterns in the figure (in particular the clustering of points near durations of 10-hours) is unclear but may arise from instrumental effects. In any case, none of our candidates have such long durations.}

\subsection{Initial Vetting}
The candidates that meet the depth and duration cutoffs (\durationthreshnumzero, \durationthreshnumone, \durationthreshnumtwo, \durationthreshnumthree, \durationthreshnumfour, and \durationthreshnumfive\ for C0-C5) were inspected by hand. Candidates were vetted in the following manner: first, obvious sinusoids and light curves that were clearly just noise were rejected, leaving 7, 18, 15, 20, 24, and 37 candidate transits or eclipsing binaries per campaign. Obvious eclipsing binaries (two transit signals of different depths, occurring about half an orbital period apart) were set aside at this point, although see \autoref{table:eb} for the EPIC number and period for likely eclipsing binaries identified throughout the search. Candidates with deep transits were rerun through EEBLS with a more permissive noise filter (50-$\sigma$) to correct our estimate of the transit depth before proceeding. A few other light curves (0, 0, 2, 2, 1, and 2, respectively) with odd patterns of photometric or stellar noise were investigated by hand to be sure they were not missed transits (none were).  The stellar parameters for all remaining planet candidates are shown in \autoref{table:stars}, with \candidateszero, \candidatesone, \candidatestwo, \candidatesthree, \candidatesfour, and \candidatesfive\ candidates, in C0-C5, respectively; \dodgycand\ candidates (all in C4) that have some questionable features are listed separately from the rest in \autoref{table:cands}.

The focus of this paper is on the shortest period planets (less than one day). However, it is common for planets with slightly longer periods to be misidentified at a shorter alias if only the shortest periods are examined. To account for this aliasing, we ran the initial EEBLS search for periods from 3-72 hours, but folded and binned the data using the period with the largest BLS spectral peak and that was less than 24 hours. This process eliminated the cases of mis-identified transiting planet candidates, although about half of the eclipsing binary signals proved to be aliases on closer examination. To ensure we had identified the correct period, we also folded all candidates' data at 0.5, 2, and 4 times the initial period. We then refined the orbital periods by running a finer search of 10,000 trial periods between 98\% and 102\% of the initial period to find the values listed in \autoref{table:cands}; see Section~\ref{section:fitting} for the derivation of the period errors.

\subsection{Follow-up observations}
\label{section:spectra}
We obtained reconnaissance spectra for 9 candidates from C1, C2, and C4 with the Tull Coud\'e spectrograph \citep{1995PASP..107..251T} at the Harlan J. Smith 2.7-m telescope at McDonald Observatory in December 2015. The exposure times ranged from 1200 to 4800 seconds, resulting in signal-to-noise ratios (S/N) from 32 to 50 per resolution element at 5650\AA. We determined stellar parameters for the host stars with the spectral fitting tool {\it Kea} \citep{2016arXiv160408170E}. We also determined absolute RVs by cross-correlating the spectra with the RV-standard star HD~50692. Our results are summarized in \autoref{table:obs}. Two observations on different nights were obtained of the relatively bright EPIC 210954046, finding large RV variations (over 20 km/s) and indicating a likely stellar rather than planetary companion.  Two other objects were found to have companions. EPIC 201606542 shows a double-peaked cross-correlation function peak, indicating an SB2 binary star; with an unknown binary period, the likely options are (a) a false positive, or (b) a  planet around one star with diluted (underestimated) planetary radius parameters due to the light of the second star. EPIC 201650711 has a fainter companion at 1\arcsec\ separation, as estimated from the guide camera, and also has likely diluted planetary parameters.

\capstartfalse
\begin{deluxetable*}{c l l l l l l c c}
\tablewidth{0pt}
\tablecaption{Candidate Planet Parameters}
\tablehead{
EPIC			&Period (d)				&$T_0$ (BJD)							&$R_p/R_*$			&$R_p$ ($R_{\oplus}$)	&$a/R_*$		&$i$~(deg)	&$\sigma_{odd-even}$	&Notes
}
\startdata
$202094740$	&$0.689647\pm0.00052$		&$2456775.19753\pm0.001$				&$0.0558_{-0.003}^{+0.005}$	&$6.58\pm0.66$	&$3.1_{-0.9}^{+0.7}$	&$78.0_{-11.1}^{+8.2}$	&$0.4$	&\\
$201264302$	&$0.212194\pm0.000026$	&$2456812.40015_{-0.00063}^{+0.00064}$	&$0.0271_{-0.002}^{+0.004}$	&$0.77\pm0.18$	&$3.6_{-0.8}^{+0.9}$	&$83.3_{-8.9}^{+4.8}$	&$0.6$	&V16\\
$201606542$	&$0.444372\pm0.000042$	&$2456817.74445_{-0.00109}^{+0.001}$		&$0.0136\pm0.002$			&$1.22\pm0.23$	&$8.3_{-3.8}^{+5.5}$	&$86.9_{-6.9}^{+2.3}$	&$0.6$	&V16,A \\
$201637175$	&$0.381087\pm0.000041$	&$2456867.13946_{-0.00032}^{+0.00031}$	&$0.0731_{-0.004}^{+0.014}$	&$4.31\pm0.98$	&$4.2_{-1.7}^{+0.6}$	&$83.0_{-13.6}^{+5.0}$	&$0.2$	&S15\\
$201650711$	&$0.259669\pm0.000041$	&$2456885.22582_{-0.00117}^{+0.00126}$	&$0.0102_{-0.001}^{+0.002}$	&$0.77\pm0.14$	&$3.1_{-0.8}^{+1.3}$	&$82.1_{-10.3}^{+5.7}$	&$0.4$	&V16\\
$203533312$	&$0.17566\pm0.000183$		&$2456933.98649\pm0.00038$				&$0.0248\pm0.001$			&$3.11\pm0.24$	&$1.7_{-0.2}^{+0.1}$	&$76.6_{-11.2}^{+9.4}$	&$0.5$	&\\
$205152172$	&$0.980414\pm0.000096$	&$2456959.30915\pm0.0014$				&$0.0219_{-0.002}^{+0.004}$	&$1.58\pm0.36$	&$5.3_{-1.9}^{+1.1}$	&$83.7_{-8.8}^{+4.7}$	&$0.7$	&V16\\
$206103150$	&$0.789693\pm0.000082$	&$2457042.1466_{-0.00162}^{+0.00159}$		&$0.0143\pm0.001$			&$1.79\pm0.13$	&$3.0_{-0.5}^{+0.3}$	&$81.8_{-7.7}^{+5.9}$	&$0.3$	&B15\\
$206151047$	&$0.358378\pm0.00006$		&$2456983.06634_{-0.00093}^{+0.00071}$	&$0.017_{-0.001}^{+0.002}$	&$1.65\pm0.18$	&$4.9\pm1.0$		&$85.4_{-5.5}^{+3.3}$	&$0.5$	&V16\\
$206169375$	&$0.367453\pm0.000039$	&$2457004.52447_{-0.00068}^{+0.00052}$	&$0.0246_{-0.001}^{+0.002}$	&$2.55\pm0.25$	&$5.2_{-1.0}^{+0.9}$	&$85.3_{-5.1}^{+3.4}$	&$0.6$	&V16\\
$206298289$	&$0.434827\pm0.000298$	&$2456987.60012_{-0.0009}^{+0.00083}$		&$0.0297_{-0.002}^{+0.004}$	&$1.62\pm0.28$	&$5.2_{-1.4}^{+1.3}$	&$84.8_{-6.2}^{+3.7}$	&$0.3$	&V16\\
$206417197$	&$0.442094\pm0.000086$	&$2457007.48414_{-0.00134}^{+0.00138}$	&$0.0138\pm0.001$			&$1.16\pm0.13$	&$3.2_{-0.7}^{+0.6}$	&$81.6_{-10.0}^{+6.1}$	&$0.1$	&V16\\
$210605073$	&$0.567055\pm0.000145$	&$2457127.43651\pm0.00056$				&$0.1047_{-0.004}^{+0.007}$	&$16.22\pm6.26$	&$5.7_{-1.3}^{+0.7}$	&$85.7_{-5.0}^{+3.1}$	&$0.5$	&B\\
$210707130$	&$0.684575\pm0.000143$	&$2457077.68807_{-0.00052}^{+0.00063}$	&$0.0181_{-0.001}^{+0.002}$	&$1.4\pm0.2$		&$6.2_{-1.6}^{+0.7}$	&$85.9_{-5.6}^{+2.9}$	&$0.0$	&\\
$210961508$	&$0.349935\pm0.000042$	&$2457087.65251_{-0.00041}^{+0.00043}$	&$0.0263_{-0.001}^{+0.003}$	&$2.18\pm0.25$	&$3.4_{-1.0}^{+0.5}$	&$81.8_{-11.0}^{+5.7}$	&$0.0$	&\\
$211357309$	&$0.46395\pm0.000118$		&$2457201.45767_{-0.00107}^{+0.00102}$	&$0.0186_{-0.002}^{+0.003}$	&$0.85\pm0.18$	&$4.7_{-1.6}^{+1.9}$	&$84.0_{-9.6}^{+4.4}$	&$0.3$	&\\
$211685045$	&$0.769057\pm0.00052$		&$2457201.38352_{-0.00186}^{+0.00189}$	&$0.035_{-0.002}^{+0.003}$	&$2.06\pm0.52$	&$3.3_{-0.7}^{+0.4}$	&$82.9_{-8.8}^{+5.0}$	&$0.4$	&\\
$211995325$	&$0.279258\pm0.00015$		&$2457154.45219_{-0.00174}^{+0.00175}$	&$0.1243_{-0.011}^{+0.021}$	&$8.41\pm3.31$	&$2.2_{-0.3}^{+0.7}$	&$79.8_{-11.7}^{+7.3}$	&$1.1$	&\\
$212150006$	&$0.898216\pm0.000135$	&$2457190.28605_{-0.00068}^{+0.00067}$	&$0.0449_{-0.002}^{+0.009}$	&$4.02\pm0.82$	&$6.3_{-3.2}^{+1.2}$	&$85.1_{-12.2}^{+3.6}$	&$0.3$	&\\
\hline
Vanishing transits\\
$211152484$	&$0.702084\pm0.000335$	&$2457091.55232_{-0.00147}^{+0.00151}$	&$0.016\pm0.001$			&$1.68\pm0.12$	&$3.0_{-0.6}^{+0.3}$	&$81.3_{-8.5}^{+6.1}$	&$0.9$	&C\\
\hline
False positives?\\
$210414957$	&$0.970016\pm0.000514$	&$2457069.9496_{-0.00049}^{+0.00048}$		&$0.0663_{-0.005}^{+0.003}$	&$6.29\pm0.61$	&$3.3_{-0.4}^{+1.2}$	&$77.4_{-3.7}^{+9.1}$	&$0.1$	&D\\
$210754505$	&$0.870775\pm0.000223$	&$2457077.07917_{-0.00057}^{+0.00054}$	&$0.0414_{-0.003}^{+0.004}$	&$3.97\pm0.47$	&$4.3_{-1.4}^{+1.9}$	&$80.0_{-8.1}^{+7.8}$	&$0.7$	&D\\
$210954046$	&$0.950356\pm0.000177$	&$2457128.33024_{-0.00113}^{+0.00111}$	&$0.0677_{-0.006}^{+0.008}$	&$6.94\pm1.08$	&$2.9_{-0.8}^{+1.3}$	&$74.3_{-9.5}^{+11.6}$	&$0.1$	&D,E
\enddata
\tablecomments{V16: first reported in \citet{2016ApJS..222...14V}. S15: the ``disintegrating planet'' first reported in \citet{2015ApJ...812..112S}. B15: first reported  \citet{2015ApJ...812L..18B}, aka WASP-47e. A: Companion star detected with McDonald observations; planetary parameters are diluted, possible false positive. B. Stellar temperature and radius from photometry (similar to F0 star). C: Variable depth/disappearing transits (possible false positive). Fit to all parameters assumes average transit depth is $R_p/R_*=0.0156\pm0.001$ and $R_p=1.7\pm0.19~R_{\oplus}$.  D:  Potential eclipsing binary from light curve variability. E: RV measured stellar-level variations.   }
\label{table:cands}
\end{deluxetable*}
\capstarttrue

\subsection{Fitting Candidate Transits}
\label{section:fitting}

\begin{figure*}
\includegraphics[width=0.8 \textwidth]{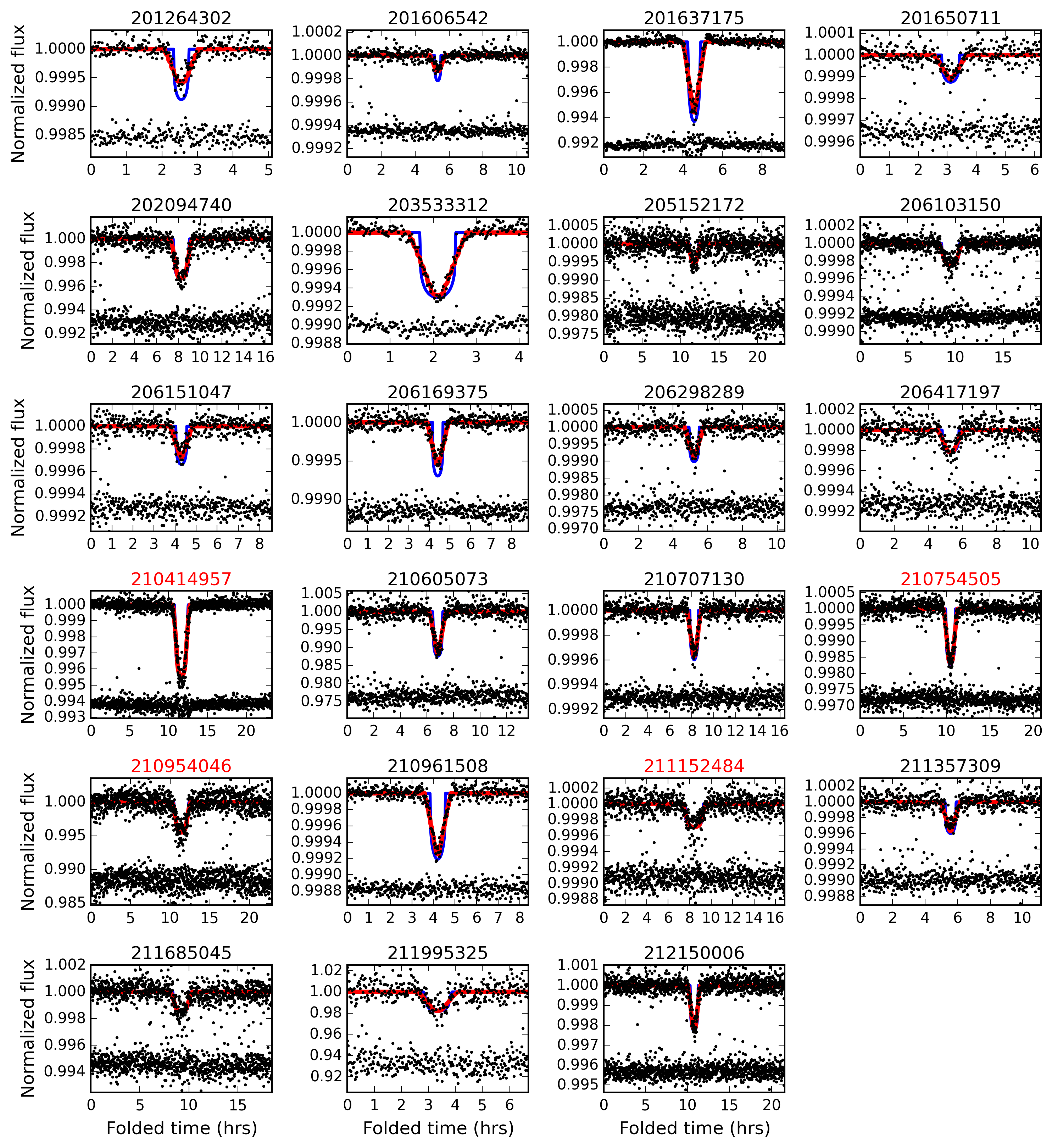}
\caption{Transiting planet-like candidates. The blue lines show the best-fit models to the planet parameters without binning over \ktwo\'s observing cadence, while the red lines are the model calculated with binning (supersampling). Residuals between the binned model fits and the data are shown below each light curve. Candidates with their names in red show signs of being eclipsing binaries or otherwise potentially non-planetary; see individual discussions in Section~\ref{section:cands}.}
\label{fig:transits}
\end{figure*}

Light curves were fit assuming a transiting planet model. The light curves were fit in Python using the algorithm from \citet{2002ApJ...580L.171M}, as implemented by the Batman package \citep{2015PASP..127.1161K}\footnote{\url{http://astro.uchicago.edu/~kreidberg/batman/}}. We used the pymodelfits\footnote{\url{https://pythonhosted.org/PyModelFit/core/pymodelfit.core.FunctionModel1D.html}} and PyMC packages to conduct a Markov Chain Monte Carlo (MCMC) transit analysis using 100,000 iterations (discarding as burn-in the first 1000 iterations and thinning the sample by a factor of 10). The results are shown in \autoref{table:cands}, with the quoted 1-$\sigma$ error bars containing 68.3\% of the posterior values. With very short transit durations, there are only a few observations per transit, resulting in considerable averaging point-to-point in the observed light curves. The solution is to fit the light curves using supersampling, where each model point is an average of several points spanning the half-hour exposure time. Since the time to run each fit scales directly with the number of points used in supersampling, we chose to use 7 points for candidates with $4\le P \le2 4$~hour periods and 11 points for the very shortest with $P<4$ hours.

Accurate stellar parameters are key to getting accurate planetary parameters. When possible, we used values for stellar $T_{eff}$,  $\log(g)$ and $[Fe/H]$ from spectral observations (either our own at McDonald or from the literature). Next in order of preference was the revised EPIC catalog \citep{2015arXiv151202643H}, which had 20 of the 23 targets of this paper (1 of which was observed from McDonald). For EPIC 202094740, we used the the K2-TESS Stellar Properties Catalog\footnote{\url{https://filtergraph.com/tess_k2campaigns}} for stellar temperature and assumed that $\log(g)=4.0$ and $[Fe/H]=0$. For EPIC 210605073, which is very faint, we estimated the stellar type as F0 from photometric \emph{ugriz} colors (see Section~\ref{sec:faintguy}) and assigned a temperature of 7020 K, as well as $\log(g)=4.0$ and $[Fe/H]=0$. 

\capstartfalse
\begin{deluxetable*}{r l rr l rrr l}
\tablecaption{Spectral Observations Of Candidates}
\startdata
\tablehead{
\colhead{EPIC} & \colhead{HJD} & \colhead{S/N} & \colhead{RV} & \colhead{$T_{\rm eff}$} & \colhead{[Fe/H]} & \colhead{$\log\left(g\right)$} & \colhead{$v\ \sin i$} & Notes \\
\colhead{} & \colhead{days} & \colhead{} & \colhead{km/s} & \colhead{K} & \colhead{} & \colhead{cm~s$^{-2}$} & \colhead{km/s}
}
201606542    & 2457462.81892  & 35    	&$0.00 \pm 0.00 $             & $5540\pm108$  &$0.030\pm 0.04$       & $4.81\pm0.12$       &$13.23\pm 0.53$ & SB2 (double peaked), A\\
203533312    & 2457462.96524  & 38   	&$18.56 \pm   1.50 $         & $6620\pm86$    &$0.090 \pm 0.05 $      & $4.19\pm0.12 $      &$23.85 \pm0.83$ &  \\
201650711    &  2457463.81728 &  43   	& $5.52 \pm   0.43 $          & $4340\pm103$ & $-0.810\pm0.08 $      & $4.25\pm 0.42$      &$ 2.31 \pm0.40$ & Comp. star at 1\arcsec\\ 
210414957    &	2457373.91896 & 32 	& $43.76 \pm0.40$		& $5838\pm 102$ & $ 0.300 \pm0.07$ 	&  $4.12 \pm0.14$ 	& $ 8.25\pm0.28$  &\\
210707130    &	2457374.67776 & 47 	& $-4.123 \pm 0.37$		& $4462\pm  84$ & $-0.280 \pm0.10$ 	&  $4.17 \pm0.18$ 	& $ 1.92\pm0.34$ &\\
210754505    &	2457374.79139 & 36 	& $-1.82 \pm0.32$		& $5875\pm 103$ & $ 0.040 \pm0.05$ 	&  $4.04 \pm0.19$ 	& $ 8.75\pm0.22$ &\\
210954046    &	2457373.65486 & 38 	& $40.57 \pm   3.91$		& $6062\pm 274$ & $ 0.000 \pm0.16$ 	&  $3.00 \pm0.58$ 	& $55.00\pm2.89$ & Large RV variation\\
--                   & 2457374.84919 & 50 	& $18.83 \pm 1.96$		& $6188\pm 298$ & $ 0.000 \pm0.16$ 	&  $3.00 \pm0.58$ 	& $53.33\pm3.55$& Large RV variation\\
210961508    &	2457375.65275 & 42	 	& $-58.53\pm 0.22$		& $4925\pm  45$ & $-0.060 \pm0.07$ 	&  $3.42 \pm0.08$ 	& $ 2.42\pm0.15$ &\\
211152484    &	2457374.82461 & 44 	& $-49.55\pm 0.23$		& $6188\pm  67$ & $-0.120 \pm0.06$ 	&  $4.12 \pm0.15$ 	& $ 8.42\pm0.15$ &
\enddata
\tablecomments{A: The parameters for 201606542 have larger uncertainties than quoted due to the presence of a secondary set of lines.  }
\label{table:obs}
\end{deluxetable*}
\capstarttrue

For EPIC 206103150, aka WASP-47, we used the published stellar radius of \citet{2013A&A...558A.106M}. For all other stars we derived stellar radii from $T_{\rm eff}$ using \citet[][their Equation 4 and Table 10]{Boyajian2012}. Where available, we compared these radius values to the EPIC values and found significant differences for only two stars: EPIC 210414957 (one of the likely false positives; EPIC $R_* = 2.319\pm 0.233~R_{\odot}$ whereas we calculated $R_* = 0.806\pm0.041~R_{\odot}$), and EPIC 210961508 (EPIC $R_* = 2.589\pm0.512~R_{\odot}$ whereas we calculated $R_* =  0.773\pm0.045~R_{\odot}$). The calculated errors on the stellar radius are included in the estimates of planetary radius. We calculated the quadratic limb darkening coefficients for the Kepler bandpass using \citet{2011A&A...529A..75C}, where we fixed microturbulent velocity = 2 km/s and used the [Fe/H] and $\log(g)$ values available except as noted previously. 

To estimate uncertainties on each candidate's orbital period, we fit for the transit ephemeris. Since there are too few points in any individual light curve to constrain a transit fit, consecutive transits were folded and binned on a number $n$ of orbital periods, with $n$ large enough to give the resulting binned transits sufficient S/N to be analyzed. (We required $n$ between 11 and 31 orbits for our candidates.) Then each folded/binned transit was fit with a linear ephemeris, and the errors on the orbital periods were assigned based on these fits. (There was no evidence of transit-timing variations for any of our candidates.) We used a similar procedure in \citet{2013ApJ...779..165J} and were able to recover the ephemerides assumed for synthetic transits injected into real data from the {\it Kepler} mission.

Since a blend scenario with similarly sized eclipsing binary stars can resemble a transiting planet with half the true period of the system, we separately fit the odd and even-numbered transits and compared the radius ratios derived for each \citep{2010ApJ...713L.103B}. The difference between the depth of the odd and even transits, expressed in terms of the errors, is shown in \autoref{table:cands} as $\sigma_{odd-even}$. All candidates with odd-even ratios greater than 3-$\sigma$ were assigned to be eclipsing binaries.

To further investigate the possibility of blend scenarios for our candidates, we looked for correlations between photometric centroids and flux variations -- in-transit shifts of the photocenter may indicate blending with objects near the target star in the sky \citep{2010ApJ...713L.103B}. For C0-2 we used the positions calculated by A. Vanderburg (private communication), while for C3-5 we used the values for MOM\_CENTR1 and MOM\_CENTR2 provided with the standard mission light curves from MAST, which we downloaded separately for each of our candidates. We found no statistically significant (3-$\sigma$) photocenter variations in the centroid position during transit compared to out of transit for any of our candidates.

\section{Candidates}
\label{section:cands}

\subsection{EPIC 206103150, a.k.a. WASP-47}

EPIC 206103150 is the WASP-47 system, where a 4-day hot Jupiter \citep{2012MNRAS.426..739H} has recently been found to be accompanied by a super-Earth at 0.8 days (the candidate identified by this survey) and another transiting planet at 9 days \citep{2015ApJ...812L..18B}, as well as a Jupiter-sized, non-transiting planet at $\sim572$~d \citep{2016A&A...586A..93N}. Clear evidence of the outer two transiting planets is seen in the folded light curve for the inner planet in \autoref{fig:transits}. This system provided a check on our detection and characterization algorithms, particularly for multiples; we saw no signs of additional planets among our other candidate systems. Our radius of $1.79\pm0.13~R_{\oplus}$ is very similar to the value of $1.829\pm0.070$ from \citet{2015ApJ...812L..18B}, which used the higher-precision short-cadence (1 min) data.




\subsection{Sub-Jovian Desert Candidates: EPIC 201637175, 202094740, 203533312, 212150006, and 211995325}

No confirmed planets with radii between about 3 and 11 $R_{\oplus}$ are known below a period of about 1.5 days, and that radius range is underpopulated out to about 3 days. This region in period-radius space has been referred to as the sub-Jovian desert \citep{2013ApJ...763...12B, 2016ApJ...820L...8M} and corresponds to the size range in which a planet would need to have significant volatiles to match the observed radius, which might be difficult for the planets to retain. One candidate for the sub-Jovian desert is, in fact, EPIC 201637175, the disintegrating planet candidate of \citet{2015ApJ...812..112S}. 

Membership in the sub-jovian desert is highly dependent on the precision with which the stellar parameters are known. EPIC 211995325 is around a very faint star ($Kp=18.2$), and the large errors on its radius ($R_{\rm p}=8.41\pm3.31 R_{\oplus}$) are due to both the low S/N light curve and the uncertain stellar size ($R_*=0.62\pm0.22 R_{\odot}$).

The remaining three taregets, EPIC 202094740, 203533312 and 212150006, are prime targets for follow-up, even though they have strong \emph{a priori} odds of being false positives \cite[see][]{2015MNRAS.452.3001C}. Many initially-promising candidates in this size range turn out to be deeper eclipses that have been diluted to appear planetary in size -- this is likely the case for the three objects discussed in Section \ref{section:dodgy}. We note that although EPIC 212150006 appears in the Kepler EB catalog online, it appears with erroneous values (see Section~\ref{section:ebs}), so its status is uncertain. Any confirmed planets would be of great value for understanding the stability and evolution of planets in this size range. The shortest-period candidate in this paper is EPIC 203533312, which is around a bright star ($Kp = 12.5$) with  $P=4.2$ hours and  $R_p=3.11\pm0.24~R_{\oplus}$, placing it at the lower edge of the sub-Jovian desert. Following the work of \citet{2013ApJ...773L..15R}, we can estimate a minimum mass for the planet by requiring that it is just exterior to its Roche limit. We find $M_p \ge 48.5~M_{\oplus}$, with a density of $\rho=8.9~{\rm g/cm^3}$.

\subsection{EPIC 210605073}
\label{sec:faintguy}

Candidate 210605073 is a 1\% deep transit around an extremely faint star ($Kp=17.9$) with no stellar parameters (e.g. $T_{\rm eff}$ and $R_{\odot}$) listed in the EPIC or TESS/Vanderbilt catalogs. Based on the broadband Sloan photometry provided on the K2 ExoFOP (\url{https://exofop.ipac.caltech.edu/k2/edit_target.php?id=210605073}), the star has very neutral to slightly blue colors ($g-r = -0.037$, $g-i=-0.145$, and $g-z=-0.266$). Using \citet[][their Table 1]{2012ApJS..199...30P} to estimate $T_{\rm eff}$ from Sloan colors, we find that the colors are just above the high-mass end of their fiducial models, suggesting that the star is a bit larger than $1.5~M_{\odot}$. We note however that no measurements of the metallicity, log g, or photometric extinction are available to better constrain the estimate. An F0 star with $M_*=1.7~M_{\odot}$ and $R_*=1.3~R_{\odot}$ implies a candidate radius of 14 $R_{\oplus}$, while (for comparison) an M-dwarf with a stellar radius of 0.4 $R_{\odot}$ implies a radius of 5 $R_{\oplus}$. Both radius values are large enough that the candidate would probably have a significant volatile component \citep{Adams2008}, and Roche-lobe overflow \citep{2015ApJ...813..101V} and/or photoevaporative mass loss \citep{2013ApJ...776....2L} would likely have removed the atmosphere of such a planet in a 13-hour orbital period. Thus, it seems \emph{a priori} unlikely this candidate is a planet, although additional (and not likely forthcoming) data would be required to make a definitive judgment.

\begin{figure}
\includegraphics[width=0.5\textwidth]{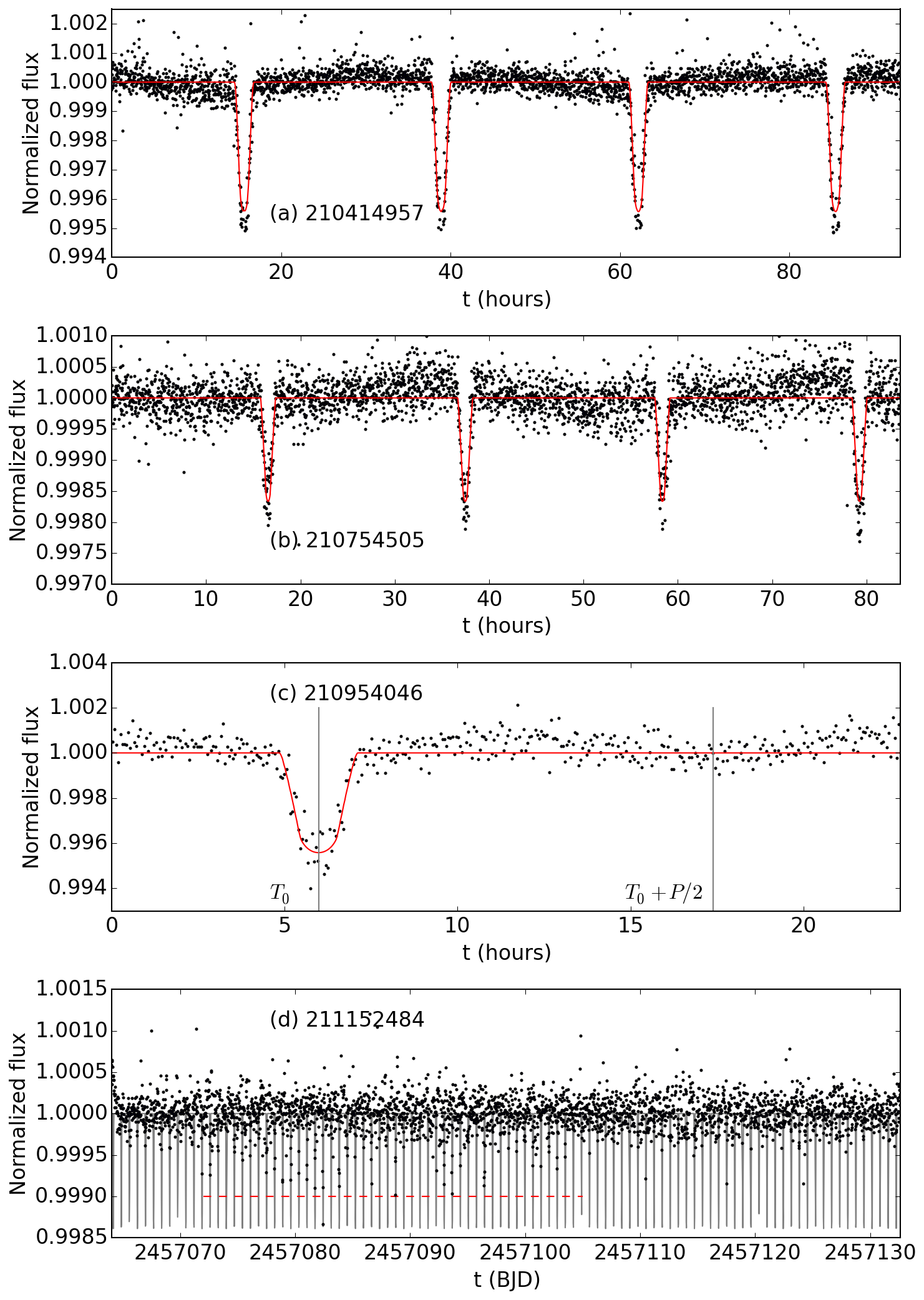}
\caption{Questionable candidates. (a)-(b)Two candidates, EPIC 21041957 and 201754505, folded to four times the orbital period, have significant out-of-transit variation and the transits occur at different points in the sinusoidal pattern; this is a common feature of similar-sized eclipsing binary light curves. (c) EPIC 210954046, folded to the orbital period, has a sinusoidal pattern with a minimum suggestive of a secondary around 0.5 orbital phase; coupled with the large RV signature, this is a confirmed false positive. (d) EPIC 211152484, showing only intermittent transit-like signals in the raw time series. The grey line shows a transit model with the parameters in \autoref{table:cands}, scaled to the depth of the deepest point (radius ratio = 0.035) which occurs around T=2457082; the red dashed horizontal line shows the approximate time range during which transit-like signals most frequently occurred.}
\label{fig:dodgy}
\end{figure}

\subsubsection{Likely EBs: out-of-transit variability for EPIC 210414957, 210754505, and 210954046}
\label{section:dodgy}

Although there are no signs of odd-even depth variations for any of the objects presented in this paper, three of them show variability in the out of transit portion of their light curves, probably explained by a false positive scenario. For EPIC 210414957 and 210754505, the out of transit portion shows a sinusoidal pattern at twice the period of the candidate transits, with transits occurring at both peaks and troughs as seen in \autoref{fig:dodgy} panels (a) and (b). Such variability is typically due to ellipsoidal variations, although for large hot planets a similar pattern could be due to phase curve variations \citep[e.g. $\upsilon$ And,][]{2006Sci...314..623H}. If the undulation were due to the phase variability of a large planet rotating through view, we would expect that the transits would only occur at same point in the variation (typically the troughs, when no emission from the planet is seen). Both objects are also in the so-called sub-Jovian desert, where stability of their atmospheres is questionable. 

The third object, EPIC 210954046, has out-of-transit variability that more closely resembles a shallow secondary eclipse (see \autoref{fig:dodgy} panel (c)). Follow-up RV observations revealed an RV shift of more than 20 km/s and indicate this candidate is a false positive (see \autoref{table:obs}); the host star is also a fast rotator. In addition, a nearby star at 2\arcsec\ separation was seen in the McDonald guide camera, lending more support to the idea that the nominal $6~R_{\oplus}$ transit is actually a diluted binary star.

\subsubsection{Variable and disappearing: EPIC 211152484}

The importance of examining the full time series, not just the binned composite transit, is illustrated by EPIC 211152484. Although the period signal in \autoref{fig:transits} strongly resembles a transiting planet, it does so only intermittently. There are no transits for about half of the period of observation (see \autoref{fig:dodgy}), and when transits are visible, they have variable depths. The best fit reported in \autoref{table:cands} is a fit to all of the data and is thus an average of transit depths; the deepest point (at $T=2457082.4$) corresponds to a radius ratio of 0.035, or double the average. Possible explanations include some kind of stellar variability, a rapidly-precessing cloud of debris, or a disintegrating comet. A full examination of this interesting object is beyond the scope of this work. Additional data may be forthcoming as part of the GO proposal from Charbonneau and colleagues to characterize and provide masses for planets smaller than 2.5~$R_{\oplus}$ (\url{http://keplerscience.arc.nasa.gov/data/k2-programs/GO4029.txt}).


\subsection{Eclipsing binaries}
\label{section:ebs}

In the process of identifying transiting planet candidates, \ebnum\ likely eclipsing binary stars were also identified (\autoref{table:eb}). Eclipsing binaries with strongly dissimilar depths were obvious on first visual inspection of the transit light curves, while those with more similar transit depths often initially were identified by EEBLS at an alias of half the true period. \autoref{table:eb} lists the EPIC number and period for all eclipsing binaries, along with a note for which were found at an alias of the true period. Because of our focus on objects with periods of under a day, the list of eclipsing binaries is not complete for $P>1$~day, and everything listed with a period of 1 day or more was originally identified at a shorter alias. It is also possible that some eclipsing binaries with periods less than one day were missed if they more closely resembled sinusoidal noise.

We cross-correlated our list of eclipsing binaries with the Kepler EB catalog (\url{http://keplerebs.villanova.edu/}), which has been extended to include K2 targets in the online Third Revision (Beta) updated 2015 Oct 26. Out of the \ebnum\ EBs listed in \autoref{table:eb}, 58 appear in that catalog (although 4 appear at half the period we identify), while 33 are new. Only two of the ultra-short-period transiting planet candidates in \autoref{table:cands} appeared in the Kepler EB catalog, and it is unclear if either truly belongs there: for EPIC 211685045, the secondary marked appears to be a glitch (3 points of noise that happen to lie half an orbital phase from the transit, see \url{http://keplerebs.villanova.edu/overview/?k=211685045}). EPIC 212150006 (which is a sub-Jovian desert candidate) has a secondary depth and width of -1 (likely an error). Nonetheless, the possibility of these and other targets being undetected eclipsing binary blends remains, and requires follow-up RV observations and high-resolution images to fully resolve.

\begin{figure}
\includegraphics[width=0.5\textwidth]{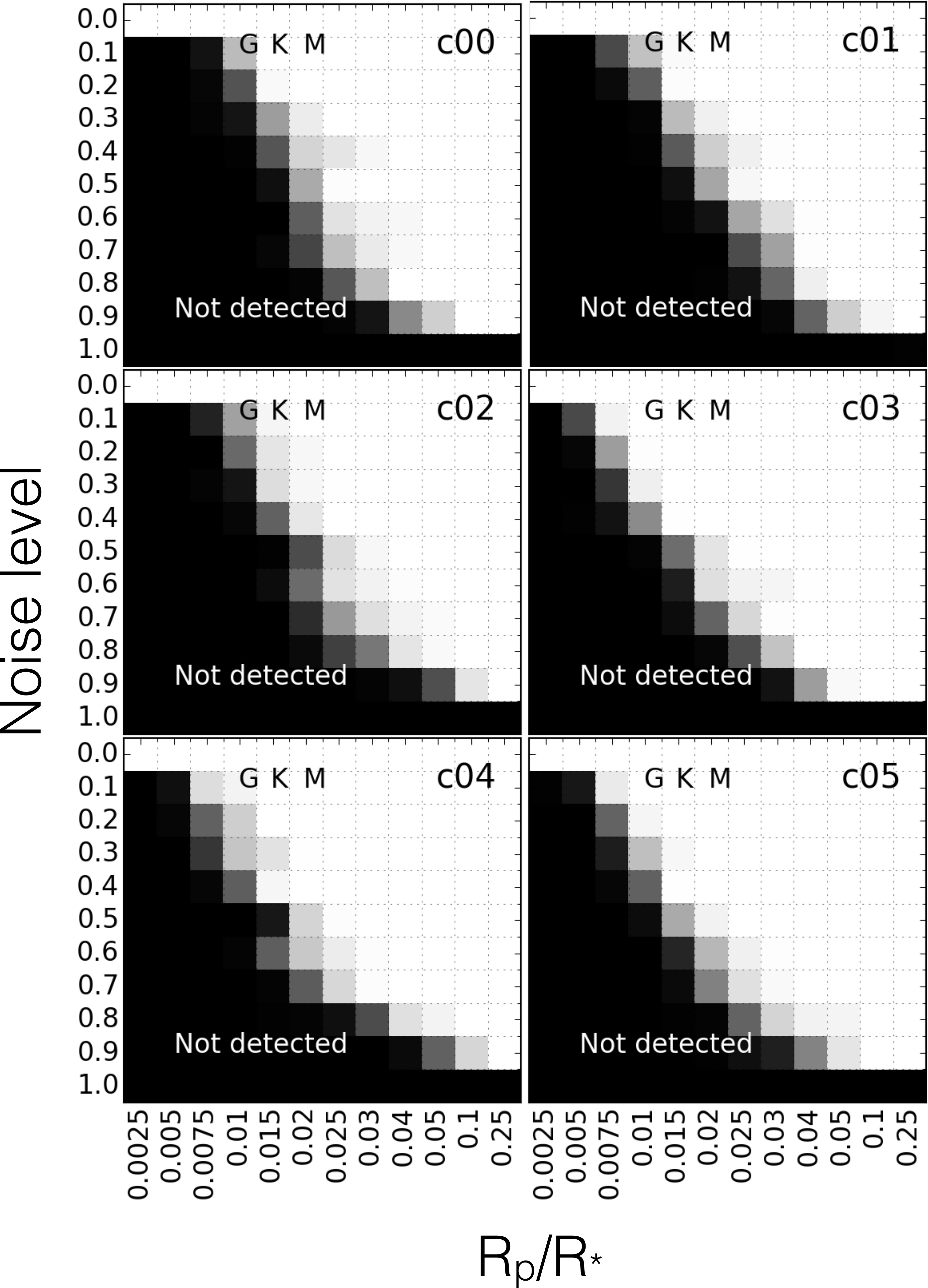}
\caption{Detectability of injected transits by radius ratio and noise of light curve. Ten light curves at the highest and lowest noise levels, and at each intervening decile, were tested for a suite of injected planets with periods from 3.5-23.5 hours and radius ratios of 0.0025 to 0.25. Black represents no detection, while the greyscale represents the fraction of samples (averaged over each of the ten light curves per point and all orbital periods) that were detected. The radius ratios of a $1R_{\oplus}$ planet around G, K, and M dwarfs are also indicated. Sensitivity improved between the first three campaigns (C0-C2) and the last three (C3-C5).}
\label{fig:completeness}
\end{figure}

\section{Survey Completeness}
\label{section:survey_completeness}

\subsection{Detectability calculations}
\begin{figure}
\includegraphics[width=0.5\textwidth]{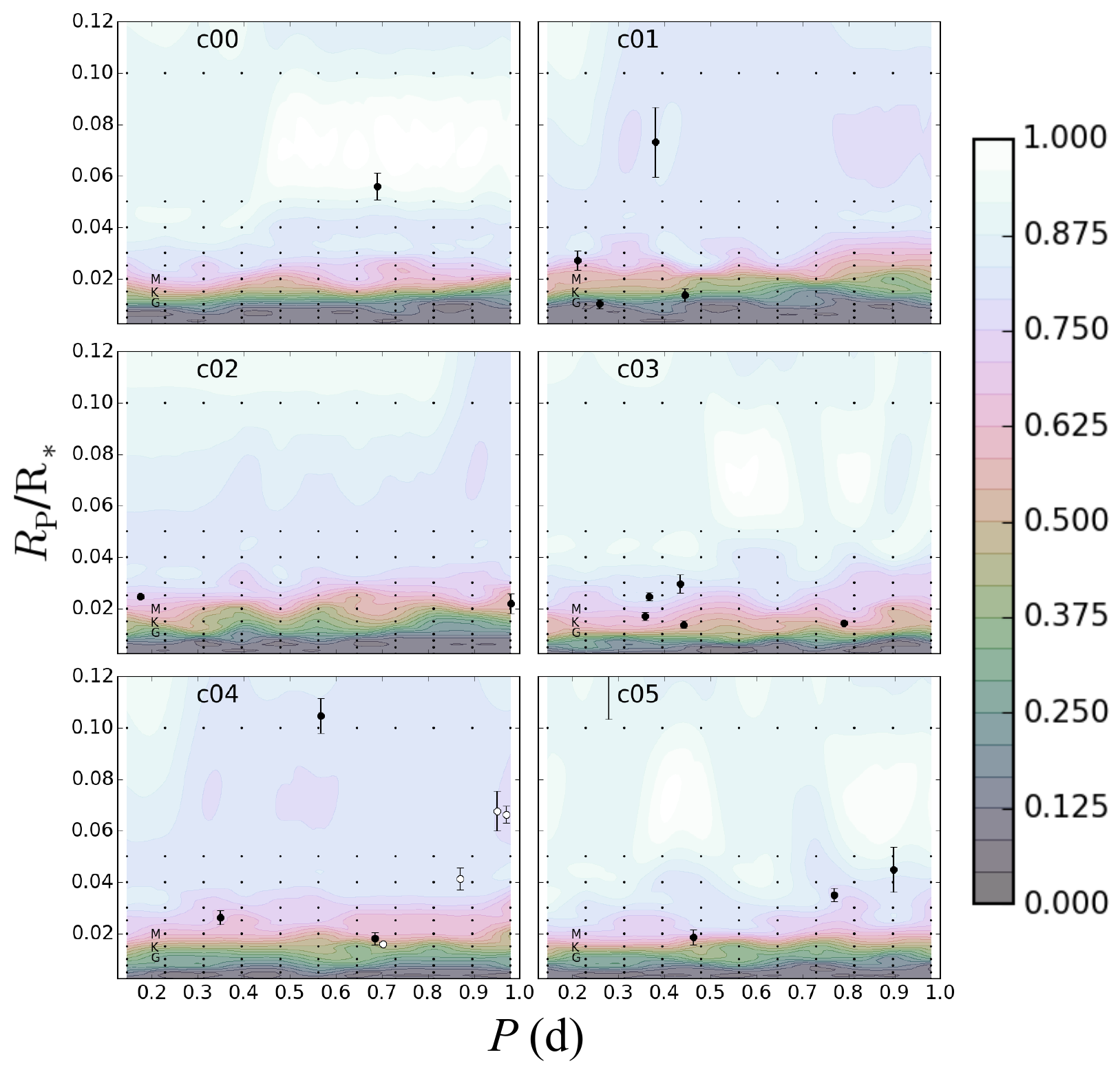}
\caption{Detectability of injected transits by radius ratio and orbital period. Ten light curves at the highest and lowest noise levels, and at each intervening decile, were tested for a suite of injected planets with periods from 3.5-23.5 hours and radius ratios of 0.0025 to 0.25. Sampled grid is shown as small black dots, while the background contours show that grid interpolated onto an evenly spaced grid. Black represents no detection, with the colorbar shading to light to represent the fraction of samples (averaged over each of the ten light curves per point and all noise levels) that were detected. The approximate location of a 1-$R_{\oplus}$ planet around G, K, and M dwarfs are also indicated (lower left). Candidates are shown with error bars (white points are likely false positives); note that one candidate from C5, 211995325, which orbits a tiny late M dwarf, is off the scale with a radius ratio of $R_p/R_*=0.1558$.}
\label{fig:period_radius}
\end{figure}

To test the completeness of our survey, we injected a grid of synthetic transits into a representative sample of light curves. We sorted all light curves from a given campaign by the median noise of the detrended light curve and took ten light curves at each decile (median noise for C0-5 respectively: 540, 1090, 700, 490, 440, and 460 ppm), as well as the least ($\approx 1$ ppm) and most ($\approx 10^6$ ppm) noisy curves for each campaign. Into each of these representative light curves we inserted transits with radius ratios from 0.0025 to 0.25 and with periods between 3.5 and 23.5 hours. Each synthetic light curve was run through our EEBLS detection pipeline, and a transit was declared detected if it was recovered within 0.1\% of the injected period and within a factor of 2 on depth. The completeness as a function of the input radius ratio and light curve noise is shown in \autoref{fig:completeness}, while the completeness as a function of period and radius is shown in \autoref{fig:period_radius}. 

As expected, detectability is a strong function of the radius ratio. Detectability also varied by campaign, with C0-C2 less sensitive to small transits than C3-C5. In C0-2 only about 10\% of light curves would have been sensitive to a 1-$R_{\oplus}$ planet around a G-star (radius ratio of 0.01), compared to 30\% for K stars and 50\% M-dwarfs. For C3-C5, the detectable fraction of Earth-sized planets is roughly 30, 50, and 65\%. We detected 3 sub-Earth-sized candidates (201264302, 201650711, and 211357309) and 7 candidates with $1\le R_p\le 2~R_{\oplus}$ (201606542, 205152172, 206103150, 206151047, 206298289, 206417197, 210707130), all around stars smaller than the Sun (\autoref{table:stars}).

By the time the radius ratio reaches 0.035, the detection probability saturates at 80-85\% of all injected transits, indicating that 15-20\% of \ktwo\ light curves are unlikely to detect transits of any size. Unlike in the analysis of \citet{SanchisOjeda2014} of the {\it Kepler} field, we did not find a strong detectability trend with orbital period, although there is a modestly increased likelihood of finding closer planets.

\subsection{Comparison to published occurrence rates}

\citet{SanchisOjeda2014} provided robust estimates of USP occurrence rate for the {\it Kepler} field, and we calculated the estimated number of candidates assuming that the same occurrence applies to C0-5 \ktwo\ targets. For this purpose, we used the effective temperatures of the \ktwo\ target stars as provided by the K2-TESS catalog, and the occurrence rates and period distribution for USPs from \citet{SanchisOjeda2014}. We also note that \citet{SanchisOjeda2014} found no candidates larger than $\sim2~R_{\oplus}$, so we have separated our candidates into two bins, using $2.2~R_{\oplus}$ as the cutoff since that is the first zero-object bin in their Figure 9.

First we estimated the expected number of planets based on the occurrence rate, which varies by stellar type. We considered all stars inferred to be dwarf stars and divided them by stellar type (based on their effective temperatures) to estimate the total number of USPs in orbit around all stars of that type. For example, \citet{SanchisOjeda2014} estimated an occurrence rate among M-dwarfs of $1.1\pm0.4$\%. In C2, there are 1,030 such stars, leading us to expect between 7 and 15 USPs orbiting M-dwarfs in that field. We generated a hypothetical population of USPs for each field and for each stellar type, with a random number distributed normally about the mean expected number and with a standard deviation given by the error bars on the occurrence rates. We assigned each hypothetical USP an orbital period between 3 and 24 hours, with a probability distribution given by that inferred for USPs in \citet{SanchisOjeda2014}. We selected as many stars of the type under study, drawing a random effective temperature, and converted that temperature to a stellar radius and mass using the empirical fits from \citet{Boyajian2012}. (This conversion involves extrapolation slightly above the range of temperatures considered in that study.) We used the stellar mass and period hosting each USP to calculate a semi-major axis and, combined with the stellar radius, a transit probability \citep{2007PASP..119..986B}. We considered that USP to transit if a random number uniformly distributed between 0 and 1 lay below the transit probability. We applied this procedure several times for each field and for each stellar type to achieve robust statistics and estimated uncertainties as the standard deviation of our resulting yields. This number is the ``Expected'' field in \autoref{table:expected_yield}, and varies from 5-14 planets per campaign. 

Next we estimated the fraction of targets that our survey would have detected (using the survey completeness calculation of Section~\ref{section:survey_completeness}) assuming the same occurrence rate as in \citet{SanchisOjeda2014}. The occurrence rate calculations were expressed in terms of planetary radius and only covered planets from $0.84 \le R_{\rm p} \le 2.2~R_{\oplus}$, while the survey completeness was in terms of radius ratio and covered a broader range of planetary radii, while not taking the stellar radius into account. So to reconcile the two we first found the fraction of planets from \citet{SanchisOjeda2014} as a function of planetary radius between $0.84 \le R_{\rm p} \le 2.2~R_{\oplus}$ (taken from their Figure 9). We used the same log bins they did, centered at 0.92, 1.09, 1.3, 1.55, and 1.84 $R_{\oplus}$; all of their bins from 2.2 $R_{\oplus}$ up were empty. We assumed (as they did) that stellar type did not affect the \emph{occurrence} rate of planets as a function of planetary radius, although the \emph{detectability} of planets would definitely depend on the stellar radius. To factor that in, we calculated the radius ratio of each radius bin if the object were around a nominal F, G, K, or M star (assumed to have radius values of 1.2, 1, 0.7, and 0.5 $R_{\odot}$, respectively). We then found a single function for the detectability of a planet as a function of radius ratio by integrating the completeness grid (\autoref{fig:period_radius}) over all periods, and taking the mean over all campaigns. Now for each stellar type we had the detectable fraction of planets in each radius bin (which ranged from 15\% of the 0.92 $R_{\oplus}$ bin around an F dwarf to 81\% of the 1.84 $R_{\oplus}$ bin around an M dwarf). We then convolved that with the fraction of F, G, K, and M stars per field to get the total fraction of planets from $0.84 \le R_p \le 2.2~R_{\oplus}$ around any stellar type that we would have found, resulting in an integrated detection rate of 37-42\% of objects in that size range depending on the field. This integrated detection rate was multiplied by the number of candidates in that size range in each campaign, resulting in the $N_{est}$ column in \autoref{table:expected_yield}, ranging from 0-11 objects per campaign.

How do the numbers compare? All of the expected values are within 3$\sigma$ of the estimated yields, although we note that the total number of objects is about half what we would expect. In particular, three campaigns (C1, C2, and C3) are within 1$\sigma$ of the expected value, while C0, C4 and C5 are low by 2-3$\sigma$ (and C0 has no detections). This raises the question of whether the lower detected numbers are intrinsic to (a) this survey, (b) the \ktwo\ mission (less likely without a mechanism for lower yields in particular campaigns), or (c) variability in planet occurrence rates with galactic location. The campaigns observed are a diverse sample: C0 is near the Galactic Anti-Center and C2 is near Galactic Center, while C1 and C3 are near the North and South Galactic Caps, respectively, and C4 and C5 are near clusters (the Pleiades/Hyades and the Beehive, respectively). The Kepler field, for comparison, is just off the galactic field in the Cygnus region along the Orion arm. 

We also note that we found 7 candidates larger than $2.2~R_{\oplus}$, while \citet{SanchisOjeda2014} did not find any in the \emph{Kepler} data (although \citet{2015ApJ...812..112S} did find one in \ktwo, the disintegrating EPIC 201637175). Some of those objects are quite likely to be false positives, and more follow-up observations are needed to confirm which, if any, are planets. For this reason, we consider it premature to calculate an occurrence rate for larger planets. We note, though, that the detectable fraction of such object is high (50-85\% around F-M stars) so this sample is more complete than the lower-radius subgroup. Thus, the fact that there are only 7 candidates larger than $2.2~R_{\oplus}$ (standing in for $N_{est}=7-14$ objects, assuming no false positives and a detection rate of 50-80\%), while there are 12 candidates smaller than $2.2~R_{\oplus}$ (standing in for $N_{est}=31$) is more evidence that the sub-Jovian desert is real.

\subsection{Comparison to other surveys}
Another test of survey completeness is to compare our list of detected objects to other surveys that examined the same data. Some of our candidates from C0-C3 were reported by \citet{2016ApJS..222...14V}, which was published while this manuscript was being prepared. Our single C0 candidate was not reported by that work. In C1, all four of our candidates were reported, along with four others well below our cutoff signal-to-noise ratio of 10 (with SNR=5-7) and two that we identify as EBS (EPIC 201182911 and 201563164).  In C2, we found one candidate (EPIC 203533312) that was not reported by \citet{2016ApJS..222...14V}, while they reported one target below our SNR cutoff. (That object, EPIC 203518244, had SNR=5.8 and also had a duration longer than our generous duration cutoff, indicating a potential non-planetary source of the transit signal.) In C3, we found no candidates that were not reported by \citet{2016ApJS..222...14V}, and the 3 candidates they report that we did not find all had SNR=6-7, again below our cutoff. As a test, we reran our C2 search with a lower cutoff of SNR=5, which resulted in no new planet recoveries but about twice as many light curves to sort through, indicating that our current cutoff is about as good as the current data and detection pipeline warrant.

\capstartfalse
\begin{deluxetable}{llll  |  ll }
\tablecaption{Ultra-short period planet yield from this survey}
\tablehead{
Field & $N_{Cand.}$ 				 & $N_{est}$ & $N_{Expected}$\tablenotemark{A}  & $N_{Cand.}$ 		& $N_{est}$\tablenotemark{B} \\
         &  ($\le 2.2~R_{\oplus}$) 	&  	&                              & ($>2.2R_{\oplus}$)	&  }
\startdata
C0 & 0	& 0		& $5 \pm 2$ 	& 1	& 1-2 \\
C1 & 3	& 7		& $9 \pm 3$ 	& 1	& 1-2\\
C2 & 1	& 5		& $ 8 \pm 3$ 	& 1	& 1-2 \\
C3 & 4	& 11		& $10 \pm 3$	& 1	& 1-2\\
C4 & 2	& 5		& $11 \pm 3$ 	& 1	& 1-2\\
C5 & 2	& 3		& $14 \pm 4$ 	&  2 	& 2-4\\
\hline
Total & 12	& 31		& $57 \pm 7$	& 7	& 7-14
\enddata
\tablecomments{A: Based on the occurrence rate for smaller planets of \citet{SanchisOjeda2014}. B: No occurrence rate is available for larger planets since \citet{SanchisOjeda2014} detected none, and some of these candidates are likely to be false positives. The estimated population assumes no false positives and a completeness of 50-80\% for larger USPs ($R_p > 2.2 R_{\oplus}$ and $P<1$~day). }
\label{table:expected_yield}
\end{deluxetable}
\capstarttrue

\section{Discussion}
\label{section:discussion}

\begin{figure*}
\includegraphics[width=\textwidth]{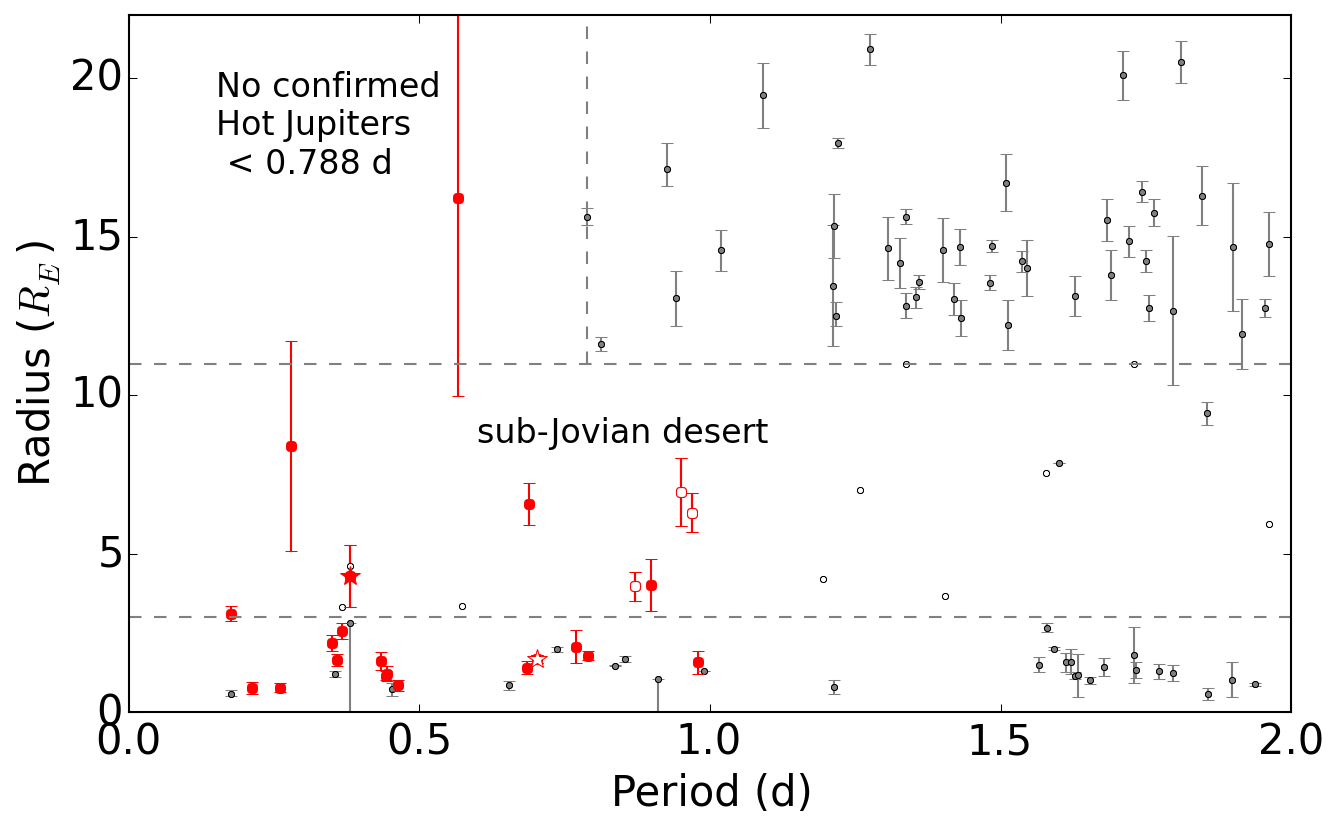}
\caption{Known exoplanets with periods under 2 days are taken from \url{http://exoplanet.eu} (queried Jan 2016) and shown as small grey circles. Candidates from the K2 ExoFOP
(\url{https://exofop.ipac.caltech.edu/k2/}, queried 2016 Apr 13) are shown as open black circles. The \totalcand\ planet candidates of this paper are shown in red (some of which were previously reported elsewhere), and the \dodgycand\ problematic objects in this paper are shown as open circles. Dashed lines indicate the regions in which few planets have been confirmed: the sub-Jovian desert, between about $3-11~R_{\oplus}$, which has no confirmed planets with $P\le1.5$~d, and Hot Jupiters with periods shorter than that of WASP-19b at $P=0.788~d$. The ``disintegrating planet'' of \citep{2015ApJ...812..112S} is shown as a filled red star just above the sub-Jovian desert line, while the object with  ``vanishing transits'' (EPIC 211152484, this work) is shown as an open red star well below the line. }
\label{fig:period-radius}
\end{figure*}

Known short-period exoplanets (less than 3 days) tend to divide into two populations: Jupiter-sized objects (10-15~$R_{\oplus}$), which are known to exist down to just under 1-day periods \citep[the shortest confirmed Hot Jupiter is WASP-19b, at 0.788 days,][]{Hebb2010}, and much smaller planets ($\le 2R_{\oplus}$) that are expected to lack volatiles. The sub-Jovian desert is the region between 3-11 $R_{\oplus}$ where there are no confirmed stable planets with periods of less than 1.5 days, as shown in \autoref{fig:period-radius}, and very few with periods less than 3-4 days. The origin of this bifurcation is an active area of research. \citet{2014ApJ...783...54K} suggest that objects with periods below a day represent represent in-migrating hot Jupiters (at the high end) and stripped remnant cores (at the low end), while \citet{2016ApJ...820L...8M} suggest the bifurcation is consistent with short-period planets originating via circularization of a high-eccentricity orbit. 

It has been noted by \citet{2015ApJ...801...41R} that most planets with $R_p >1.6~R_{\oplus}$ are not (entirely) rocky, and must therefore contain a substantial volatile fraction. We will note that in general for planets with radius values of 2-3~$R_{\oplus}$ ``volatiles'' does not always imply H/He gas. For example, Kepler-22b \citep{Borucki2012} has a radius of 2.4~$R_{\oplus}$, with its mass constrained to be below 52.8 $M_{\oplus}$ \citep{2013ApJ...777..134K}, and may be an ocean-like world. It's not clear, though, how long such an ocean would last if Kepler-22b were moved from its current orbital period of nearly 290 days into the sub-day period regime of the planets that are the focus of this work. Nor is it likely that an object with $R>3~R_{\oplus}$ could achieve that size without a substantial H/He envelope \citep{2007ApJ...659.1661F}, which would be vulnerable to tidal stripping and/or photoevaporation.

Six of our candidates have radius values of $>3~R_{\oplus}$ (202094740, 201637175, 203533312, 210605073, 211995325, and 212150006). It is entirely possible that some or perhaps all are false positives. The dearth of objects in this size range detected in other surveys makes this group highly important to pursue via follow-up observations.

We stress that our candidates are still preliminary, and require accurate followup observations to be certain that (a) we have the correct stellar parameters, since the planetary parameters scale directly with the stellar radius value, and (b) that we can rule out false positive scenarios. Towards goal (a) we are continuing efforts to obtain spectra at McDonald to determine better measurements of $T_{\rm eff}$ and $R_*$ (for most objects). For those candidate that are bright enough, we are also pursuing (b) by acquiring radial velocity and high resolution imaging to rule out false positive scenarios, and to determine the fraction of starlight that might be due to nearby stars (which will affect the measured planet radius).

\section{Conclusion}

We have conducted a survey for ultra short-period candidates (less than 1 day) in \ktwo\ data. In this paper, we present \totalcand\ candidates in C0-5 of \ktwo. Four additional objects (three likely eclipsing binaries and one odd case of vanishing transits) are also noted, as are \ebnum\ eclipsing binary systems. Among the new candidates reported are  EPIC 203533312, one of the shortest period planet candidates identified ($P=4.2$~hours), which by stability arguments must have a density of at least  $\rho$ = 8.9 g/cm$^3$.  Its star, at $Kp = 12.5$, is a good candidate for follow-up observations. It is also one of five candidates with radius values in the sub-Jovian desert between 3-11 $R_{\oplus}$ and less than $P$=1.5 days. An additional candidate Hot Jupiter (EPIC 210605073) at $P$=13.6 hours, would be the shortest-period hot Jupiter identified, although the faintness of its host star ($Kp = 17.89$) makes additional follow-up difficult. We estimated our survey completeness, which varied from 10-85\% depending on planet size, stellar type, and \ktwo\ campaign, with an average detection efficiency of about 40\% for Earths and super-Earths ($0.8 \le R_p \le 2.2~R_{\oplus}$). Finally, we compared the occurrence rate of Earths and super-Earths with $R_p \le 2.2 R_{\oplus}$ to that of \citep{SanchisOjeda2014} and find that we detected about half as many planets in \ktwo\ as they found in \emph{Kepler} (31 vs. 57 after adjusting for completeness). We note that that survey did not find any candidates with  $R_p \ge 2.2 R_{\oplus}$, while we found 8, indicating that (a) several of our candidates may yet prove to be false positives, and (b) larger planets are indeed rare among ultra-short period planets.

\acknowledgments

All the data analyzed in this paper were collected by the \ktwo\ mission, funding for which is provided by the NASA Science Mission Directorate. The data were obtained from the Mikulski Archive for Space Telescopes (MAST). STScI is operated by the Association of Universities for Research in Astronomy, Inc., under NASA contract NAS5-26555. Support for MAST for non-HST data is provided by the NASA Office of Space Science via grant NNX09AF08G and by other grants and contracts. This study is based upon work supported by NASA under Grant No. NNX15AB78G issued through the Astrophysical Data Analysis Program by Science Mission Directorate. This work has made use of the K2-TESS Stellar Properties Catalog on the Filtergraph data portal, through the TESS Science Office's target selection working group (architects K. Stassun, J. Pepper, N. De Lee, M. Paegert). The Filtergraph data portal system is trademarked by Vanderbilt University. This research used Uncertainties: a Python package for calculations with uncertainties, by Eric O. Lebigot, \url{http://pythonhosted.org/uncertainties/}. We thank Andrew Vanderburg, Daniel Huber, and an anonymous referee for helpful comments on a draft of this manuscript.

{\it Facilities:} \facility{\ktwo}.



\bibliography{k2_superpig_search}


\end{document}